\pdfoutput=1
\PassOptionsToPackage{table, dvipsnames}{xcolor}
\documentclass[sigconf]{acmart} 

\usepackage{amsmath}
\usepackage{physics}
\usepackage{dsfont}
\usepackage{mathrsfs}
\usepackage{wasysym}
\usepackage{adjustbox}

\usepackage{lineno,hyperref}
\usepackage{float}
\usepackage{multirow}
\usepackage{xcolor,colortbl}
\usepackage{adjustbox}
\usepackage{booktabs}

\usepackage{graphicx,lipsum,afterpage,subcaption}
\usepackage{enumitem}
\usepackage{xspace}
\usepackage{float}
\usepackage{tabularx} 

\expandafter\let\csname c@tblerows\endcsname\rownum

\usepackage{makecell}
\usepackage{boldline}
\usepackage{cleveref}
\usepackage{stackengine}

\usepackage[flushleft]{threeparttable}

\usepackage{bm}
\usepackage{multirow}
\usepackage{graphicx}

\usepackage{algorithm}
\usepackage[noend]{algpseudocode}

\usepackage[htt]{hyphenat}

\usepackage[frozencache=true,cachedir=.,newfloat]{minted}

\usepackage{todonotes}

\usepackage[subtle]{savetrees}

\definecolor{White}{gray}{0.995} \usepackage{colortbl}

\usepackage[font=small,skip=1pt]{caption}
\usepackage{listings}

\SetupFloatingEnvironment{listing}{name=Configuration}

\newcommand{\myrowcolour}{\rowcolor[gray]{0.925}}

\newcommand{\alejandro}[1]{\textcolor{blue}{{\bf [Alejandro: }{\em #1}{\bf ]}}} 

\newcommand{\claudio}[1]{\textcolor{orange}{{\bf [Claudio: }{#1}{\bf ]}}}

\newcommand{\felice}[1]{\textcolor{violet}{{\bf [Felice: }{#1}{\bf ]}}}

\def\framework{\textsc{Elliot}\xspace}
\def\elliot{\textsc{Elliot}\xspace}

\AtBeginDocument{%
  \providecommand\BibTeX{{%
    \normalfont B\kern-0.5em{\scshape i\kern-0.25em b}\kern-0.8em\TeX}}}

\copyrightyear{2021}
\acmYear{2021}
\setcopyright{rightsretained}
\acmConference[SIGIR '21] {Proceedings of the 44th International ACM SIGIR Conference on Research and Development in Information Retrieval}{July 11--15, 2021}{Virtual Event, Canada.}
\acmBooktitle{Proceedings of the 44th International ACM SIGIR Conference on Research and Development in Information Retrieval (SIGIR '21), July 11--15, 2021, Virtual Event, Canada}
\acmPrice{}
\acmISBN{978-1-4503-8037-9/21/07}
\acmDOI{10.1145/3404835.3463245}

\begin{document}
\fancyhead{}
\title{\framework: A Comprehensive and Rigorous Framework for Reproducible Recommender Systems Evaluation}

\settopmatter{authorsperrow=4}

\author{Vito Walter Anelli}
 \authornote{Corresponding authors: vitowalter.anelli@poliba.it, claudio.pomo@poliba.it.}
\email{vitowalter.anelli@poliba.it}
\affiliation{\institution{Politecnico di Bari, Italy}
\country{}
 }
 
 \author{Alejandro Bellogín}
\email{alejandro.bellogin@uam.es}
 \affiliation{
 \institution{Autónoma Madrid, Spain}
 \country{}
 }
 
\author{Antonio Ferrara}
\email{antonio.ferrara@poliba.it}
\affiliation{\institution{Politecnico di Bari, Italy}
\country{}
 }
\author{Daniele Malitesta}
\email{daniele.malitesta@poliba.it}
\affiliation{\institution{Politecnico di Bari, Italy}
\country{}
 }
\author{Felice Antonio Merra}
\email{felice.merra@poliba.it}
\affiliation{\institution{Politecnico di Bari, Italy}
\country{}
 }
\author{Claudio Pomo}
\authornotemark[1]
\email{claudio.pomo@poliba.it}
\affiliation{\institution{Politecnico di Bari, Italy}
\country{}
 }
 
 \author{\mbox{Francesco Maria Donini}}
 \email{donini@unitus.it}
\affiliation{
 \institution{Università della Tuscia, Italy}
\country{}
 }
 
\author{Tommaso Di Noia}
\email{tommaso.dinoia@poliba.it}
\affiliation{\institution{Politecnico di Bari, Italy}
\country{}
 }

\renewcommand{\shortauthors}{Anelli et al.}

\begin{abstract}
Recommender Systems have shown to be an effective way to alleviate the over-choice problem and provide accurate and tailored recommendations.
However, the impressive number of proposed recommendation algorithms, splitting strategies, evaluation protocols, metrics, and tasks, has made rigorous experimental evaluation particularly challenging.
Puzzled and frustrated by the continuous recreation of appropriate evaluation benchmarks, experimental pipelines, hyperparameter optimization, and evaluation procedures, we have developed an exhaustive framework to address such needs.
\framework is a comprehensive recommendation framework that aims to run and reproduce an entire experimental pipeline by processing a simple configuration file. 
The framework loads, filters, and splits the data considering a vast set of strategies ($13$ splitting methods and $8$ filtering approaches, from temporal training-test splitting to nested K-folds Cross-Validation).
\framework\footnote{\url{https://github.com/sisinflab/elliot}} optimizes hyperparameters ($51$ strategies) for several recommendation algorithms ($50$), selects the best models, compares them with the baselines providing intra-model statistics, computes metrics ($36$) spanning from accuracy to beyond-accuracy, bias, and fairness, and conducts statistical analysis (Wilcoxon and Paired t-test).\footnote{An extended version of this paper is available at \url{https://arxiv.org/abs/2103.02590}}
\end{abstract}
\iftrue 

\fi

\maketitle
\vspace{-1em}
\section{Introduction}~\label{sec:intro}
In the last decade, Recommendation Systems (RSs) have gained momentum as the pivotal choice for personalized decision-support systems.
Recommendation is essentially a retrieval task where a catalog of items is ranked and the top-scoring items are presented to the user~\cite{DBLP:conf/kdd/KricheneR20}.
Once it was demonstrated their ability to provide personalized items to clients, both Academia and Industry devoted their attention to RSs~\cite{DBLP:conf/recsys/AnelliDSSABBGKL20,bennett2007netflix, DBLP:journals/umuai/LudewigMLJ21, DBLP:conf/recsys/LudewigMLJ19}.
This collective effort resulted in an impressive number of recommendation algorithms, ranging from memory-based~\cite{DBLP:conf/www/SarwarKKR01} to latent factor-based~\cite{DBLP:conf/recsys/CremonesiKT10,funk2006netflix,DBLP:reference/sp/KorenB15,DBLP:conf/icdm/Rendle10}, and deep learning-based methods~\cite{DBLP:conf/www/LiangKHJ18,DBLP:conf/wsdm/WuDZE16}.
At the same time, the RS research community 
became conscious that \textit{accuracy} was not sufficient to guarantee user satisfaction~\cite{DBLP:conf/chi/McNeeRK06}.
\textit{Novelty} and \textit{diversity}~\cite{DBLP:conf/sigir/Vargas14,DBLP:reference/sp/CastellsHV15,DBLP:journals/toit/HurleyZ11} came into play as new dimensions to be analyzed when comparing algorithms.
However, this was only the first step in the direction of a more comprehensive evaluation of RSs.
Indeed, more recently, the presence of \textit{biased}~\cite{DBLP:conf/recsys/Baeza-Yates20,Zhu21} and \textit{unfair}~\cite{DBLP:conf/sigir/EkstrandBD19,DBLP:conf/recsys/EkstrandBD19,deldjoo2020flexible} recommendations towards user groups and item categories has been widely investigated. 
In fact, RSs have been widely studied and applied in various domains and tasks, with different (and often contradicting in their hypotheses) splitting preprocessing strategies~\cite{DBLP:journals/umuai/CamposDC14} fitting the specific scenario.
Moreover, machine learning (and recently also deep learning) techniques are prominent in algorithmic research and require their hyperparameter optimization strategies and procedures~\cite{DBLP:conf/recsys/AnelliNSPR19,DBLP:conf/recsys/ValcarceBPC18}.

The abundance of possible choices generated much confusion about choosing the correct baselines, conducting the hyperparameter optimization and the experimental evaluation~\cite{DBLP:conf/recsys/SaidB14,DBLP:conf/recsys/SaidB14a}, and reporting the details of the adopted procedure.
Consequently, two major concerns arose: unreproducible evaluation and unfair comparisons~\cite{DBLP:conf/recsys/SunY00Q0G20}. 
On the one hand, the negative effect of unfair comparisons is that various proposed recommendation models have been compared with suboptimal baselines~\cite{DBLP:conf/recsys/DacremaCJ19,DBLP:journals/corr/abs-1905-01395}.
On the other hand, in a recent study~\cite{DBLP:conf/recsys/DacremaCJ19}, it has been shown that only one-third of the published experimental results are, in fact, reproducible. 
Progressively, the RS community has welcomed the emergence of recommendation, evaluation, and even hyperparameter tuning frameworks~\cite{DBLP:conf/recsys/SunY00Q0G20,DBLP:conf/cikm/Ekstrand20,DBLP:conf/recsys/GantnerRFS11,DBLP:conf/sigir/Vargas14,DBLP:conf/icml/BergstraYC13}.
%
However,facilitating reproducibility or extending the provided functionality would typically depend on developing bash scripts or programming on whatever language each framework is written.

This work introduces \framework, a novel kind of recommendation framework, to overcome these obstacles.
The framework analyzes the recommendation problem from the researcher's perspective.
Indeed, \framework conducts a whole experiment, from dataset loading to results gathering.
The core idea is to feed the system with a simple and straightforward configuration file that drives the framework through the experimental setting choices.
\framework natively provides for widespread research evaluation features, like the analysis of multiple cut-offs and several RSs ($50$).
According to the recommendation model, the framework allows, to date, the choice among $27$ similarities, the definition of multiple neural architectures, and $51$ hyperparameter tuning combined approaches, unleashing the full potential of the HyperOpt library~\cite{DBLP:conf/icml/BergstraYC13}.
To enable the evaluation for the diverse tasks and domains, \framework supplies 36 metrics (including Accuracy, Error-based, Coverage, Novelty, Diversity, Bias, and Fairness metrics),  13 splitting strategies, and 8 prefiltering policies. 
The framework can also measure to what extent the RS results are significantly different from each other, providing the paired t-test and Wilcoxon statistical hypothesis tests. 
Finally, \framework lets the researcher quickly build their models and include them in the experiment.

\vspace{-1em}
\iffalse
\section{Background}~\label{sec:background}
\else
\section{Prior work}
\noindent \textbf{Background.} 
\fi
RS evaluation is an active, ever-growing research topic related to reproducibility, which is a cornerstone of the scientific process as identified by~\citet{DBLP:conf/recsys/KonstanA13}.
Recently researchers have taken a closer look at this problem, in particular because depending on how well we evaluate and assess the efficacy of a system, the significance and impact of such results will increase.

Some researchers argue that to enhance reproducibility, and to facilitate fair comparisons between different works (either frameworks, research papers, or published artifacts), at least the following four stages must be identified within the evaluation protocol~\cite{DBLP:conf/recsys/SaidB14}:
\textit{data splitting}, \textit{item recommendations}, \textit{candidate item generation}, and \textit{performance measurement}.
In a recent work~\cite{DBLP:journals/corr/abs-2102-00482}, these stages have been completed with \textit{dataset collection} and \textit{statistical testing}.
Some of these stages can be further categorized, such as performance measurement, depending on the performance dimension to be analyzed (e.g., ranking vs error, accuracy vs diversity, and so on).

In fact, the importance and relevance of the aforementioned stages have been validated in recent works; however, even though most of the RS literature has been focused on the impact of the item recommendation stage as an isolated component, this is far from being the only driver that affects RS performance or the only component impacting on its potential for reproducibility.
In particular, \citet{DBLP:conf/recsys/MengMMO20} survey recent works in the area and conclude that no standard splitting strategy exists, in terms of random vs temporal splits; furthermore, the authors found that the selection of the splitting strategy can have a strong impact on the results.
Previously, \citet{DBLP:journals/umuai/CamposDC14} categorized and experimented with several variations of random and temporal splitting strategies, evidencing the same inconsistency in the results.
Regarding the candidate item generation, it was first shown~\cite{DBLP:conf/recsys/BelloginCC11} that different strategies selecting the candidate items to be ranked by the recommendation algorithm may produce results that are orders of magnitude away from each other; this was later confirmed~\cite{DBLP:conf/recsys/SaidB14} in the context of benchmarking recommendation frameworks.
Recent works~\cite{DBLP:conf/kdd/KricheneR20,DBLP:conf/kdd/LiJGL20} evidenced that some of these strategies selecting the candidate items may introduce inconsistent measurements which should, hence, not be trusted.

Finally, depending on the recommendation task and main goal of the RS, several performance dimensions, sometimes contradicting,  can be assessed. For a classical overview of these dimensions, we refer the reader to \citet{DBLP:reference/sp/GunawardanaS15}, where metrics accounting for prediction accuracy, coverage, confidence, trust, novelty, diversity, serendipity, and so on are defined and compared. However, to the best of our knowledge, there is no public implementation  providing more than one or two of these dimensions.
Moreover, recently the community has considered additional dimensions such as bias (in particular, popularity bias~\cite{DBLP:conf/aies/Abdollahpouri19}) and fairness~\cite{DBLP:conf/sigir/EkstrandBD19}.
These dimensions are gaining attention, and several metrics addressing different subtleties are being proposed, but no clear winner or standard definition emerged so far -- as a consequence, the community lacks an established implementation of these novel evaluation dimensions.


\iffalse
\section{Related Frameworks}~\label{sec:related}
\else
\noindent \textbf{Related Frameworks.} 
\fi
Reproducibility is the keystone of modern RSs research. \citet{DBLP:conf/recsys/DacremaCJ19} and~\citet{DBLP:conf/recsys/RendleKZA20} have recently raised the need of comprehensive and fair recommender model evaluation. Their argument on the outperforming recommendation accuracy of latent-factor models over deep-neural ones, when  an extensive hyper-parameter tuning was performed, made it essential the development of novel recommendation frameworks. 
Starting from 2011, Mymedialite~\cite{DBLP:conf/recsys/GantnerRFS11}, LensKit~\cite{DBLP:conf/cikm/Ekstrand20,DBLP:conf/recsys/EkstrandLKR11}, LightFM~\cite{DBLP:conf/recsys/Kula15}, RankSys~\cite{DBLP:conf/sigir/Vargas14}, and Surprise~\cite{DBLP:journals/jossw/Hug20}, have formed the basic software for rapid prototyping and testing of recommendation models, thanks to an easy-to-use model execution and the implementation of standard accuracy, and beyond-accuracy, evaluation measures and splitting techniques. However, the outstanding success and the community interests in deep learning (DL) recommendation models, raised need for novel instruments. LibRec~\cite{DBLP:conf/um/GuoZSY15}, Spotlight~\cite{kula2017spotlight}, and OpenRec~\cite{DBLP:conf/wsdm/YangBGHE18} are the first open-source projects that made DL-based recommenders available -- with less than a dozen of available models without filtering, splitting, and hyper-optimization tuning strategies. An important step towards more exhaustive and up-to-date set of model implementations have been released with RecQ~\cite{DBLP:conf/icdm/Yu0YLGW19}, DeepRec~\cite{DBLP:conf/isca/GuptaHSWRWL0W20}, and Cornac~\cite{DBLP:journals/jmlr/SalahTL20} frameworks. However, they do not provide a general tool for extensive experiments on the pre-elaboration and the evaluation of a dataset. Indeed, after the reproducibility hype~\cite{DBLP:conf/recsys/DacremaCJ19,DBLP:conf/recsys/RendleKZA20}, DaisyRec~\cite{DBLP:conf/recsys/SunY00Q0G20} and RecBole~\cite{DBLP:journals/corr/abs-2011-01731} raised the bar of framework capabilities, making available both large set of models, data filtering/splitting operations and, above all, hyper-parameter tuning features. However, we found a significant gap in splitting and filtering capabilities, in addition to a complete lack of two nowadays popular (even critical) aspects of recommendation performance: biases and fairness. 
Reviewing these related frameworks, emerged a striking lack of an open-source recommendation framework able to perform \emph{by design} an extensive set of pre-elaboration operations, to support several hyperparameters optimization strategies and multiple sets 
of evaluation measures, which include bias and fairness ones, supported by statistical significance tests -- a feature absent in other frameworks (as of February 2021). \framework meets all these needs.
Table \ref{table:framework} gives an overview of the frameworks and to which extent they satisfy the mentioned requirements. 



\begin{figure*}[!t]
    \centering
    \includegraphics[height=0.3\textheight]{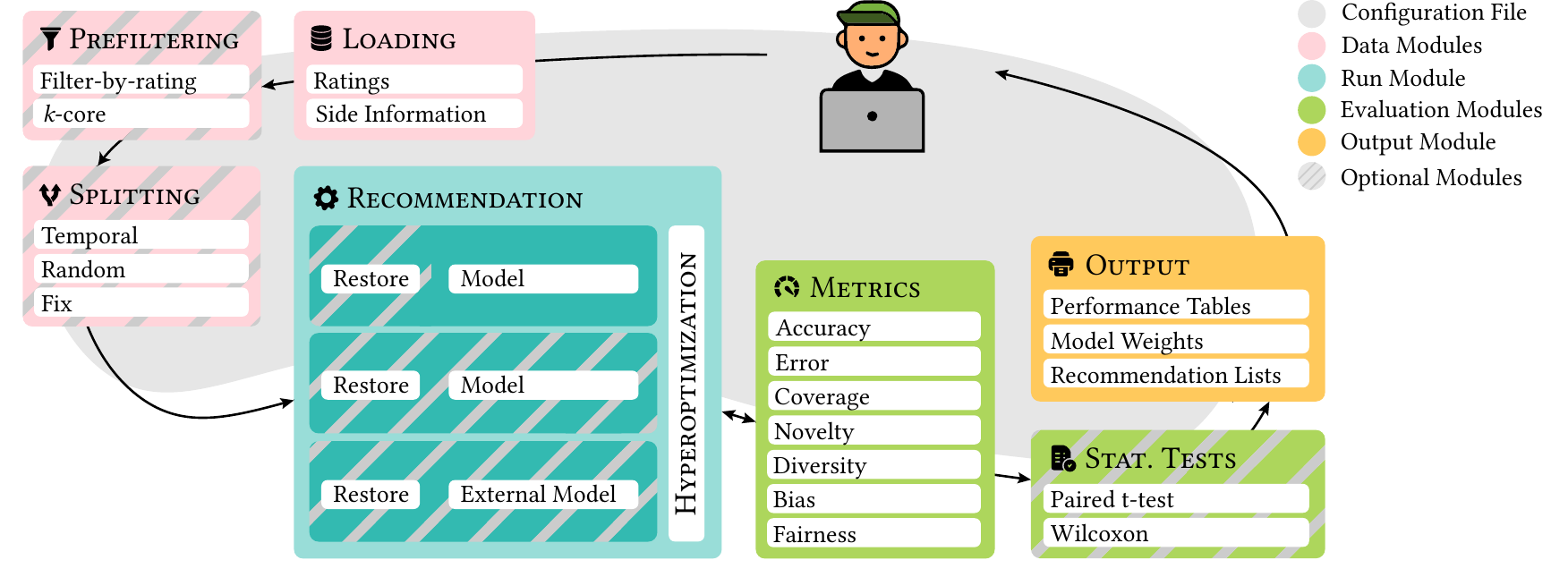}
    \caption{Overview of \elliot.}
    \label{fig:framework}
   \vspace{-1em}
\end{figure*}
\vspace{-1em}
\section{\framework}~\label{sec:framework}
\framework is an extensible framework composed of eight functional modules, each of them responsible for a specific step of an experimental recommendation process. What happens under the hood (\Cref{fig:framework}) is transparent to the user, who is only expected to provide human-level experimental flow details using a customizable configuration file. Accordingly, \framework builds the overall pipeline. The following sections deepen into the details of the eight \framework's modules and outline the preparation of a configuration file.


\vspace{-1em}
\subsection{Data Preparation}~\label{sec:framework_data}
The \textit{Data} modules are responsible for handling and managing the experiment input, supporting various additional information, e.g., item features, visual embeddings, and images. After being loaded by the \textit{Loading} module, the input data is taken over by \textit{Prefiltering} and \textit{Splitting} modules whose strategies are reported in~\Cref{table:framework}.
\textbf{\subsubsection{Loading}~\label{sec:framework_data_loading}}
RSs experiments could require different data sources such as user-item feedback or additional side information, e.g., the visual features of an item images.
To fulfill these requirements,
\framework comes with different implementations of the \textit{Loading} module.
Additionally, the user can design computationally expensive prefiltering and splitting procedures that can be stored and loaded to save future computation. 
Data-driven extensions can handle additional data like visual features~\cite{DBLP:conf/icdm/KangFWM17,DBLP:conf/sigir/ChenZ0NLC17}, and semantic features extracted from knowledge graphs~\cite{DBLP:conf/semweb/AnelliNSRT19}.
Once a side-information-aware \textit{Loading} module is chosen, it filters out the items devoiding the required information to grant a fair comparison.
\textbf{\subsubsection{Prefiltering}~\label{sec:framework_data_filtering}}
After data loading, \elliot provides data filtering operations through two possible strategies.
The first strategy implemented in the \textit{Prefiltering} module is \textit{Filter-by-rating}, which drops off a user-item interaction if the preference score is smaller than a given threshold.
It can be (i) a \textit{Numerical} value, e.g., $3.5$, (ii) a \textit{Distributional} detail, e.g., global rating average value, or (iii) a user-based distributional (\textit{User Dist.}) value, e.g., user's average rating value.
The second prefiltering strategy, $k$\textit{-core}, filters out users, items, or both, with less than $k$ recorded interactions. 
The $k$\textit{-core} strategy can proceed iteratively (\textit{Iterative} $k$\textit{-core}) on both users and items until the $k$\textit{-core} filtering condition is met, i.e., all the users and items have at least $k$ recorded interaction. Since reaching such condition might be intractable, \elliot allows specifying the maximum number of iterations (\textit{Iter-$n$-rounds}). Finally, the \textit{Cold-Users} filtering feature allows retaining cold-users only.
\textbf{\subsubsection{Splitting}~\label{sec:framework_data_splitting}}
If needed, the data is served to the \textit{Splitting} module. In detail, \elliot provides (i) \textit{Temporal}, (ii) \textit{Random}, and (iii) \textit{Fix} strategies. The \textit{Temporal} strategy splits the user-item interactions based on the transaction timestamp, i.e., fixing the timestamp, finding the optimal one~\cite{DBLP:conf/ecir/AnelliNSRT19,DBLP:conf/recsys/BelloginS17}, or adopting a hold-out (\textit{HO}) mechanism. 
The \textit{Random} strategy includes hold-out (\textit{HO}), $K$-repeated hold-out (\textit{K-HO}), and cross-validation (\textit{CV}). \Cref{table:framework} provides further configuration details. Finally, the \textit{Fix} strategy exploits a precomputed splitting.

\subsection{Recommendation Models}~\label{sec:framework_model}
After data loading and pre-elaborations,  \textit{Recommendation} module (\Cref{fig:framework}) provides the functionalities to train (and restore) the \elliot recommendation models and the new ones integrated by users.

\noindent \textbf{\subsubsection{Implemented Models}~\label{sec:framework_model_rec}}
\elliot integrates, to date, $50$ recommendation models (see \Cref{table:framework}) partitioned into two sets. The first set includes $38$ \textit{popular} models implemented in at least two of frameworks reviewed in this work (i.e., adopting a framework-wise popularity notion).  
\Cref{table:framework} shows that \elliot is the framework covering the largest number of popular models, with $30$ models out of $38$, i.e., $79\%$. 
The second set comprises other well-known state-of-the-art recommendation models implemented in less than two frameworks,
namely, BPRSLIM~\cite{DBLP:conf/icdm/NingK11}, ConvNCF~\cite{DBLP:conf/ijcai/0001DWTTC18}, NPR~\cite{DBLP:conf/wsdm/NiuCL18}, MultiDAE~\cite{DBLP:conf/www/LiangKHJ18}, and NAIS~\cite{DBLP:journals/tkde/HeHSLJC18}, 
graph-learning based, i.e., NGCF~\cite{DBLP:conf/sigir/Wang0WFC19}, and LightGCN~\cite{DBLP:conf/sigir/0001DWLZ020},
visual-based, i.e., VBPR~\cite{DBLP:conf/aaai/HeM16}, DeepStyle~\cite{DBLP:conf/sigir/LiuWW17}, DVBPR~\cite{DBLP:conf/icdm/KangFWM17}, ACF~\cite{DBLP:conf/sigir/ChenZ0NLC17}, and VNPR~\cite{DBLP:conf/wsdm/NiuCL18},
adversarial-robust, i.e., APR~\cite{DBLP:conf/sigir/0001HDC18} and AMR~\cite{DBLP:journals/tkde/TangDHYTC20}, 
generative adversarial network (GAN)-based, i.e., IRGAN~\cite{DBLP:conf/sigir/WangYZGXWZZ17} and CFGAN~\cite{DBLP:conf/cikm/ChaeKKL18},
content-aware, i.e., Attribute-I-$k$NN and -U-$k$NN~\cite{DBLP:conf/recsys/GantnerRFS11}, VSM~\cite{9143460,DBLP:conf/i-semantics/NoiaMORZ12}, Wide \& Deep~\cite{DBLP:conf/recsys/Cheng0HSCAACCIA16}, and KaHFM~\cite{DBLP:conf/semweb/AnelliNSRT19} 
recommenders.

\begin{table*}[p]
\caption{Overview of the \elliot and related frameworks functionalities.}
\small
{
\label{ref:tab:unique-a}
\renewcommand{\arraystretch}{0.9}
\setlength{\tabcolsep}{0.29em}
\rowcolors{21}{white}{gray!15}
\begin{tabularx}{\textwidth}{X@{\hskip 0.8em}llllllllr@{\hskip 0.8em}llllllllllllllllllr@{\hskip 0.8em}llllllllr}

\hlineB{2.5}
\myrowcolour
 \multicolumn{ 38}{c}{\textbf{DATA ELABORATION AND MODEL OPTIMIZATION STRATEGIES}} \\

\toprule

 & \multicolumn{ 9}{c}{\textbf{Prefiltering}}
  & \multicolumn{ 19}{c}{\textbf{Splitting}}

 & \multicolumn{ 9}{c}{\multirow{1}{*}{\textbf{Hyperparam.} \textbf{Tuning} }} 
\\
 \cmidrule(r){2-10}\cmidrule(lr){11-29}\cmidrule(l){30-38}

 & \multicolumn{ 3}{c}{\multirow{2}{*}{\begin{tabular}[c]{@{}c@{}} Filter-\\by-rating \end{tabular}}} 
 & \multicolumn{ 5}{c}{\multirow{2}{*}{$k$-core}} 
 & \multicolumn{1}{l}{} 
  & \multicolumn{ 7}{c}{Temporal} 
 & \multicolumn{ 10}{c}{Random} 
 & \multicolumn{ 1}{c}{\multirow{2}{*}{Fix}} 
 & \multicolumn{ 1}{c}{} 
 & \multicolumn{ 4}{c}{\multirow{2}{*}{\begin{tabular}[c]{@{}c@{}} Search \\ Strategy\end{tabular}}}
 & \multicolumn{ 4}{c}{\multirow{2}{*}{\begin{tabular}[c]{@{}c@{}} Search \\\ Space\end{tabular}}}
  & \multicolumn{1}{l}{}
 \\
 \cmidrule(l){11-17}\cmidrule(lr){18-27}

  & \multicolumn{ 3}{l}{} 
 & \multicolumn{ 5}{l}{} 
  & \multicolumn{1}{l}{} 
 & \multicolumn{ 2}{c}{\textit{Fix}} 
 & \multicolumn{ 5}{c}{\textit{HO}} 
 & \multicolumn{ 5}{c}{\textit{HO}} 
 & \multicolumn{ 3}{c}{$K$\textit{-HO}} 
 & \multicolumn{ 2}{c}{\textit{CV}} 
 \\ 
  \cmidrule(r){2-4}\cmidrule(lr){5-9}
  \cmidrule(lr){11-12} \cmidrule(lr){13-17}
  \cmidrule(lr){18-22} \cmidrule{23-25} 
  \cmidrule(lr){26-27}  \cmidrule(lr){28-28}    
  \cmidrule(r){30-33} \cmidrule(l){34-37}

 & \rotatebox{90}{Numerical}
 & \rotatebox{90}{Distributional}
 & \rotatebox{90}{User Dist.}
  & \rotatebox{90}{User}
 & \rotatebox{90}{Item}
 & \rotatebox{90}{Iterative}
 & \rotatebox{90}{Iter-$n$-rounds}
 & \multicolumn{1}{c}{\rotatebox{90}{Cold-Users}}
 & \rotatebox{90}{\textbf{Tot.}}

 & \rotatebox{90}{Best timestamp} 
 & \rotatebox{90}{Handcrafted}
 & \rotatebox{90}{By-Ratio Sys.}
 & \rotatebox{90}{By-Ratio User}
 & \rotatebox{90}{Leave-$1$-out}
 & \rotatebox{90}{Leave-$n$-in}
 & \rotatebox{90}{Leave-$n$-out}
 & \rotatebox{90}{By-Ratio Sys.}
 & \rotatebox{90}{By-Ratio User}
 & \rotatebox{90}{Leave-$1$-out}
 & \rotatebox{90}{Leave-$n$-in}
 & \rotatebox{90}{Leave-$n$-out}
 & \rotatebox{90}{By-Ratio Sys.}
 & \rotatebox{90}{Leave-$1$-out}
 & \rotatebox{90}{Leave-$n$-out}
 & \rotatebox{90}{$k$-folds Sys.}
 & \rotatebox{90}{$k$-folds User}
 & \multicolumn{1}{c}{\rotatebox{90}{Computed}}
 & \rotatebox{90}{\textbf{Tot.}} 

 & \rotatebox{90}{Grid}
 & \rotatebox{90}{Annealing}
 & \rotatebox{90}{Bayesian} 
 & \rotatebox{90}{Random}
 & \rotatebox{90}{Fix}
 & \rotatebox{90}{Uniform}
 & \rotatebox{90}{Normal}
 & \multicolumn{1}{c}{\rotatebox{90}{Log-Uniform}}
 & \rotatebox{90}{\textbf{Tot.}}
 
\\ 
\toprule

LensKit Java~\cite{DBLP:conf/recsys/EkstrandLKR11}   &  &  &  &  &  &  &  &  & 0   &  &  &  & \checkmark & \checkmark & \checkmark & \checkmark &  & \checkmark & \checkmark & \checkmark & \checkmark &  &  &  &  & \checkmark & \checkmark & 10 &  &  &  &  &  &  &  &  & 0 \\ 
Mymedialite~\cite{DBLP:conf/recsys/GantnerRFS11}   &  &  &  &  &  &  &  &  & 0 &  &  & \checkmark & \checkmark &  &  & \checkmark & \checkmark &  &  &  &  &  &  &  & \checkmark &  & \checkmark & 6  &  &  &  &  &  &  &  &  & 0 \\ 
RankSys~\cite{DBLP:conf/sigir/Vargas14}  &  &  &  &  &  &  &  &  & 0       &  &  &  &  &  &  &  &  &  &  &  &  &  &  &  &  &  & \checkmark & 1  &  &  &  &  &  &  &  &  & 0 \\ 
LibRec~\cite{DBLP:conf/um/GuoZSY15}   &  &  &  &  &  &  &  &  & 0       &  &  & \checkmark & \checkmark & \checkmark & \checkmark &  & \checkmark & \checkmark & \checkmark & \checkmark &  &  & \checkmark &  & \checkmark &  & \checkmark & 11  &  &  &  &  &  &  &  &  & 0 \\ 
Implicit~\cite{frederickson2018fast}  &  &  &  &  &  &  &  &  & 0   &  &  &  &  &  &  &  &  &  &  &  &  &  &  &  &  &  & \checkmark & 1  &  &  &  &  &  &  &  &  & 0 \\ 
OpenRec~\cite{DBLP:conf/wsdm/YangBGHE18}   &  &  &  &  &  &  &  &  & 0     &  &  &  &  &  &  &  &  &  &  &  &  &  &  &  &  &  & \checkmark & 1   &  &  &  &  &  &  &  &  & 0 \\ 
RecQ~\cite{DBLP:conf/icdm/Yu0YLGW19}  &  &  &  &  &  &  &  &  & 0 &  &  &  &  &  &  &  &  & \checkmark &  &  &  & \checkmark &  &  &  &  & \checkmark & 3  &  &  &  &  &  &  &  &  & 0 \\ 
DeepRec~\cite{DBLP:conf/isca/GuptaHSWRWL0W20} &  &  &  &  &  &  &  &  & 0 &  &  &  &  &  &  &  & \checkmark &  &  &  &  &  &  &  &  &  & \checkmark & 2   &  &  &  &  &  &  &  &  & 0 \\ 
LensKit Python~\cite{DBLP:conf/cikm/Ekstrand20} &  &  &  &  &  &  &  &  & 0 &  &  &  &  &  &  &  &  &  &  &  &  &  & \checkmark & \checkmark & \checkmark & \checkmark & \checkmark & 5   &  &  &  &  &  &  &  &  & 0 \\ 
Surprise~\cite{DBLP:journals/jossw/Hug20}   &  &  &  &  &  &  &  &  & 0  &  &  &  &  &  &  &  &  & \checkmark & \checkmark &  &  & \checkmark & \checkmark &  & \checkmark &  & \checkmark & 6  & \checkmark &  &  & \checkmark & \checkmark &  &  &  & 3 \\ 
Cornac~\cite{DBLP:journals/jmlr/SalahTL20} & \checkmark &  &  & \checkmark & \checkmark &  &  &  & 3  &  &  &  &  &  &  &  & \checkmark & \checkmark &  &  &  &  &  &  & \checkmark &  & \checkmark & 4   & \checkmark &  &  & \checkmark & \checkmark & \checkmark &  &  & 4 \\ 
RecBole~\cite{DBLP:journals/corr/abs-2011-01731}  &  &  &  &  &  &  &  &  & 0  &  &  &  & \checkmark & \checkmark &  & \checkmark & \checkmark & \checkmark & \checkmark &  & \checkmark &  &  &  &  &  & \checkmark & 8  & \checkmark & \checkmark & \checkmark & \checkmark & \checkmark & \checkmark &  & \checkmark & 7 \\ 
DaisyRec~\cite{DBLP:conf/recsys/SunY00Q0G20} & \checkmark &  &  & \checkmark & \checkmark &  &  &  & 3  &  &  & \checkmark & \checkmark & \checkmark &  &  & \checkmark & \checkmark & \checkmark &  &  &  &  &  &  &  & \checkmark & 7  &  &  & \checkmark &  & \checkmark & \checkmark &  & \checkmark & 4 \\ 


\framework & \textbf{\checkmark} & \textbf{\checkmark} & \textbf{\checkmark} & \textbf{\checkmark} & \textbf{\checkmark} & \textbf{\checkmark} & \textbf{\checkmark} & \textbf{\checkmark} & \textbf{8} & \textbf{\checkmark} & \textbf{\checkmark} & \textbf{} & \textbf{\checkmark} & \textbf{\checkmark} & \textbf{} & \textbf{\checkmark} & \textbf{} & \textbf{\checkmark} & \textbf{\checkmark} & \textbf{} & \textbf{\checkmark} & \textbf{\checkmark} & \textbf{\checkmark} & \textbf{\checkmark} & \textbf{} & \textbf{\checkmark} & \textbf{\checkmark} & \textbf{13}  & \textbf{\checkmark} & \textbf{\checkmark} & \textbf{\checkmark} & \textbf{\checkmark} & \textbf{\checkmark} & \textbf{\checkmark} & \textbf{\checkmark} & \textbf{\checkmark} & \textbf{8} \\ 

\bottomrule
\end{tabularx}
}
\vspace{0.7em}

{
\label{ref:tab:unique-b}
\renewcommand{\arraystretch}{0.9}
\setlength{\tabcolsep}{0.205em}
\rowcolors{6}{white}{gray!15}
\begin{tabularx}{\textwidth}{X@{\hskip 0.6em}llllllllllllllllllllllllllllllllllllll@{\hskip 0.6em}r@{\hskip 0.6em}r}

\hlineB{2.5}
\myrowcolour \multicolumn{ 41}{c}{\textbf{RECOMMENDER MODELS}}
\\
\toprule
 & \multicolumn{ 39}{c}{\textbf{Popular Models}}
\\
 \cmidrule(r){2-39}

 & \rotatebox{90}{MostPop} 
 & \rotatebox{90}{Random} 
 & \rotatebox{90}{I-$k$NN\cite{DBLP:journals/internet/LindenSY03}} 
 & \rotatebox{90}{U-$k$NN\cite{DBLP:conf/cscw/ResnickISBR94}} 
 & \rotatebox{90}{PureSVD\cite{DBLP:reference/sp/KorenB15}} 
 & \rotatebox{90}{FunkSVD\cite{funk2006netflix}} 
 & \rotatebox{90}{SVD$++$\cite{DBLP:conf/kdd/Koren08}} 
 & \rotatebox{90}{SLIM\cite{DBLP:conf/icdm/NingK11}} 
 & \rotatebox{90}{MF\cite{DBLP:journals/computer/KorenBV09}} 
 & \rotatebox{90}{WRMF\cite{DBLP:conf/sdm/ZhangWFM06}} 
 & \rotatebox{90}{FM\cite{DBLP:conf/icdm/Rendle10}} 
 & \rotatebox{90}{NeuMF\cite{DBLP:conf/www/HeLZNHC17}} 
 & \rotatebox{90}{BPRMF\cite{DBLP:conf/uai/RendleFGS09}} 
 & \rotatebox{90}{DMF\cite{DBLP:conf/ijcai/XueDZHC17}}
 & \rotatebox{90}{FISM\cite{DBLP:conf/kdd/KabburNK13}} 
 & \rotatebox{90}{NNMF\cite{DBLP:journals/tii/LuoZXZ14}} 
 & \rotatebox{90}{NFM\cite{DBLP:conf/sigir/0001C17}} 
 & \rotatebox{90}{SoREC\cite{DBLP:conf/cikm/MaYLK08a}} 
 & \rotatebox{90}{DeepFM\cite{DBLP:conf/ijcai/GuoTYLH17}} 
 & \rotatebox{90}{PMF\cite{DBLP:conf/nips/SalakhutdinovM07}} 
 & \rotatebox{90}{AFM\cite{DBLP:conf/ijcai/XiaoY0ZWC17}} 
 & \rotatebox{90}{FFM\cite{DBLP:conf/recsys/JuanZCL16}} 
 & \rotatebox{90}{WBPR\cite{DBLP:journals/jmlr/GantnerDFS12}} 
 & \rotatebox{90}{DSSM\cite{DBLP:conf/cikm/HuangHGDAH13}} 
 & \rotatebox{90}{VBPR\cite{DBLP:conf/aaai/HeM16}} 
 & \rotatebox{90}{ConvMF\cite{DBLP:conf/recsys/KimPOLY16}} 
 & \rotatebox{90}{GMF\cite{DBLP:conf/www/HeLZNHC17}} 
 & \rotatebox{90}{Caser\cite{DBLP:conf/wsdm/TangW18}} 
 & \rotatebox{90}{MultiVAE\cite{DBLP:conf/www/LiangKHJ18}} 
 & \rotatebox{90}{I-AutoR\cite{DBLP:conf/www/SedhainMSX15}} 
 & \rotatebox{90}{U-AutoR\cite{DBLP:conf/www/SedhainMSX15}} 
 & \rotatebox{90}{CDAE\cite{DBLP:conf/wsdm/WuDZE16}} 
 & \rotatebox{90}{CML\cite{DBLP:conf/www/HsiehYCLBE17}} 
 & \rotatebox{90}{LogMF\cite{johnson2014logistic}} 
 & \rotatebox{90}{LDA\cite{DBLP:conf/recsys/KrestelFN09}} 
 & \rotatebox{90}{SoMF\cite{DBLP:conf/recsys/JamaliE10}} 
 & \rotatebox{90}{SlopeOne\cite{DBLP:conf/sdm/LemireM05}} 
 & \rotatebox{90}{SoReg\cite{DBLP:conf/wsdm/MaZLLK11}}  
 & \rotatebox{90}{\textbf{Others}} 
 & \rotatebox{90}{\textbf{Tot.}}\\ \toprule

LensKit Java~\cite{DBLP:conf/recsys/EkstrandLKR11}  &  &  & \checkmark & \checkmark &  & \checkmark &  &  & \checkmark &  &  &  &  &  &  &  &  &  &  &  &  &  &  &  &  &  &  &  &  &  &  &  &  &  &  &  & \checkmark &  & 0 & 5 \\

Mymedialite~\cite{DBLP:conf/recsys/GantnerRFS11} & \checkmark & \checkmark & \checkmark & \checkmark &  &  & \checkmark & \checkmark & \checkmark & \checkmark &  &  & \checkmark &  &  &  &  &  &  &  &  &  &  &  &  &  &  &  &  &  &  &  &  & \checkmark &  & \checkmark & \checkmark &  & 6 & 18 \\

RankSys~\cite{DBLP:conf/sigir/Vargas14} & \checkmark & \checkmark & \checkmark & \checkmark &  &  &  &  & \checkmark &  & \checkmark &  &  &  &  &  &  &  &  & \checkmark &  &  &  &  &  &  &  &  &  &  &  &  &  &  & \checkmark &  &  &  & 0 & 8 \\

LibRec~\cite{DBLP:conf/um/GuoZSY15} & \checkmark & \checkmark & \checkmark & \checkmark &  &  & \checkmark & \checkmark & \checkmark & \checkmark & \checkmark &  & \checkmark &  & \checkmark & \checkmark & \checkmark & \checkmark &  & \checkmark &  & \checkmark & \checkmark &  &  & \checkmark &  &  &  & \checkmark & \checkmark & \checkmark &  &  & \checkmark & \checkmark & \checkmark & \checkmark & 30 & 55 \\

Implicit~\cite{frederickson2018fast} &  &  & \checkmark & \checkmark &  &  &  &  &  &  &  &  & \checkmark &  &  &  &  &  &  &  &  &  &  &  &  &  &  &  &  &  &  &  &  & \checkmark &  &  &  &  & 0 & 4 \\

OpenRec~\cite{DBLP:conf/wsdm/YangBGHE18} &  &  &  &  &  &  &  &  &  &  &  &  & \checkmark &  &  &  &  &  &  & \checkmark &  &  &  &  & \checkmark &  &  &  &  &  &  &  & \checkmark &  &  &  &  &  & 7 & 11 \\

RecQ~\cite{DBLP:conf/icdm/Yu0YLGW19} &  &  &  &  & \checkmark &  & \checkmark &  &  &  &  &  &  &  &  &  &  & \checkmark &  & \checkmark &  &  &  &  &  &  &  &  &  &  &  &  &  &  &  & \checkmark & \checkmark & \checkmark & 6 & 13 \\

DeepRec~\cite{DBLP:conf/isca/GuptaHSWRWL0W20} &  &  &  &  &  &  &  &  & \checkmark &  & \checkmark & \checkmark & \checkmark & \checkmark &  & \checkmark & \checkmark &  &  &  & \checkmark &  &  & \checkmark &  &  & \checkmark & \checkmark &  & \checkmark & \checkmark & \checkmark & \checkmark &  &  &  &  &  & 5 & 20 \\

LensKit Python~\cite{DBLP:conf/cikm/Ekstrand20} & \checkmark &  & \checkmark & \checkmark &  & \checkmark &  &  & \checkmark &  &  &  & \checkmark &  &  &  &  &  &  &  &  &  &  &  &  &  &  &  &  &  &  &  &  &  &  &  &  &  & 0 & 6 \\

Surprise~\cite{DBLP:journals/jossw/Hug20} &  &  & \checkmark & \checkmark & \checkmark &  & \checkmark &  &  &  &  &  &  &  &  & \checkmark &  &  &  &  &  &  &  &  &  &  &  &  &  &  &  &  &  &  &  &  & \checkmark &  & 1 & 7 \\ 

Cornac~\cite{DBLP:journals/jmlr/SalahTL20} & \checkmark &  & \checkmark & \checkmark & \checkmark &  &  &  & \checkmark &  &  & \checkmark &  &  &  & \checkmark &  & \checkmark &  & \checkmark &  &  & \checkmark &  & \checkmark & \checkmark & \checkmark &  & \checkmark &  &  &  &  &  &  &  &  &  & 21 & 35 \\ 

RecBole~\cite{DBLP:journals/corr/abs-2011-01731} & \checkmark &  & \checkmark &  &  &  &  &  &  &  & \checkmark & \checkmark & \checkmark & \checkmark & \checkmark &  & \checkmark &  & \checkmark &  & \checkmark & \checkmark &  & \checkmark &  &  &  & \checkmark & \checkmark &  &  & \checkmark &  &  &  &  &  &  & 50 & \textbf{65} \\ 

DaisyRec~\cite{DBLP:conf/recsys/SunY00Q0G20} & \checkmark &  & \checkmark & \checkmark & \checkmark &  &  & \checkmark & \checkmark & \checkmark & \checkmark & \checkmark &  &  &  &  & \checkmark &  & \checkmark &  & \checkmark &  &  &  &  &  &  &  & \checkmark &  &  & \checkmark &  &  &  &  &  &  & 0 & 14 \\

\framework & \checkmark & \checkmark & \checkmark & \checkmark & \checkmark & \checkmark & \checkmark & \checkmark & \checkmark & \checkmark & \checkmark & \checkmark & \checkmark & \checkmark & \checkmark & \checkmark & \checkmark &  & \checkmark & \checkmark & \checkmark & \checkmark &  &  & \checkmark & \checkmark & \checkmark &  & \checkmark & \checkmark & \checkmark & & \checkmark & \checkmark &  &  & \checkmark &  & 20 & 50 \\

\bottomrule
\end{tabularx}

}

\vspace{0.7em}

{
\label{ref:tab:unique-c}
\setlength{\tabcolsep}{0.8em}
\renewcommand{\arraystretch}{0.9}
\rowcolors{7}{gray!15}{white}
\begin{tabularx}{\textwidth}{X@{\hskip 1.8em}rrlllllr@{\hskip 1.8em}ccr}

\hlineB{2.5}
\myrowcolour \multicolumn{ 12}{c}{\textbf{EVALUATION OF RECOMMENDATION PERFORMANCE}}
\\

\toprule
 & \multicolumn{8}{c}{\textbf{Metric Families}} & \multicolumn{3}{c}{\textbf{Statistical Tests}} \\ 
 \cmidrule(r){2-9}\cmidrule(l){10-12}
& Accuracy 
& Error
& Coverage 
& Novelty 
& Diversity 
& Bias 
& Fairness 
& \textbf{Tot.}
& Paired t-test 
& Wilcoxon
& \textbf{Tot.} \\ 

\toprule

LensKit Java~\cite{DBLP:conf/recsys/EkstrandLKR11}      & 2 & 2 & \multicolumn{1}{r}{1} &  & \multicolumn{1}{r}{1} &  &  & 6 &&&0\\
Mymedialite~\cite{DBLP:conf/recsys/GantnerRFS11}        & 6 & 2 &  &  &  &  &  & 8 && &0\\ 
RankSys~\cite{DBLP:conf/sigir/Vargas14}                 & 7 & & \multicolumn{1}{r}{\textbf{3}} & \multicolumn{1}{r}{\textbf{6}} & \multicolumn{1}{r}{\textbf{18}} &  &  & 34 && &0\\ 
LibRec~\cite{DBLP:conf/um/GuoZSY15}                         & 6 & \textbf{4} &  &  &  &  &  & 10 && &0\\ 
Implicit~\cite{frederickson2018fast} & & &  &  &  &  &  & 0  &&&0\\ 
OpenRec~\cite{DBLP:conf/wsdm/YangBGHE18}                        & 4 & &  &  &  &  &  & 4 &&&0\\ 
RecQ~\cite{DBLP:conf/icdm/Yu0YLGW19}                            & 6 & 2 &  &  &  &  &  & 8 && &0\\ 
DeepRec~\cite{DBLP:conf/isca/GuptaHSWRWL0W20}               & 6 & 2 &  &  &  &  &  & 8 && &0\\ 
LensKit Python~\cite{DBLP:conf/cikm/Ekstrand20}                         & 4 & 2 &  &  &  &  &  & 6 && &0\\ 
Surprise~\cite{DBLP:journals/jossw/Hug20} & & \textbf{4} &  &  &  &  &  & 4 && &0\\ 
Cornac~\cite{DBLP:journals/jmlr/SalahTL20} & 8 & 3 &  &  &  &  &  & 11 && &0\\ 
RecBole~\cite{DBLP:journals/corr/abs-2011-01731} & 8 & 3 &  &  &  &  &  & 11 &&&0\\ 
DaisyRec~\cite{DBLP:conf/recsys/SunY00Q0G20} & 8 & &  &  &  &  &  & 8 &&&0\\ 
\framework & \textbf{11} & 3 & \multicolumn{1}{r}{\textbf{3}} & \multicolumn{1}{r}{2} & \multicolumn{1}{r}{3} & \multicolumn{1}{r}{\textbf{10}} & \multicolumn{1}{r}{\textbf{4}} & \textbf{36} & \checkmark & \checkmark &\textbf{2}\\ 
\bottomrule
\end{tabularx}

}

\label{table:framework}
\end{table*}

\vspace{-1em}
\textbf{\subsubsection{Hyper-parameter Tuning}~\label{sec:framework_model_hyperopt}}
Hyperparameter tuning is an ingredient of the recommendation model training that definitely influences its performance~\cite{DBLP:conf/recsys/RendleKZA20}. 
\elliot provides \textit{Grid Search}, \textit{Simulated Annealing}, \textit{Bayesian Optimization}, and \textit{Random Search} strategies. 
Furthermore, \elliot allows performing four traversing strategies across the search space defined in each recommendation model configuration. 
When the user 
details the possible hyperparameters (as a list) without specifying a search strategy, \elliot automatically performs an exhaustive \textit{Grid Search}. 
\elliot may exploit the full potential of the \textit{HyperOpt}~\cite{DBLP:conf/icml/BergstraYC13} library by considering all its sampling strategies. 
\Cref{table:framework} summarizes the available \textit{Search Strategies} and \textit{Search Spaces}.


\subsection{Performance Evaluation}~\label{sec:framework_result}
After the training phase, \elliot continues its operations, evaluating recommendations. 
\Cref{fig:framework} indicates this phase with two distinct evaluation modules: Metrics and Statistical Tests.
\textbf{\subsubsection{Metrics}~\label{sec:framework_result_metrics}}
\elliot provides $36$ evaluation metrics (see~\Cref{table:framework}), partitioned into seven families: \textit{Accuracy}~\cite{DBLP:conf/kdd/ZhouZSFZMYJLG18,Schroder201178}, \textit{Error}, \textit{Coverage}, \textit{Novelty}~\cite{DBLP:conf/recsys/VargasC11}, \textit{Diversity}~\cite{DBLP:conf/sigir/ZhaiCL03}, \textit{Bias}~\cite{DBLP:conf/flairs/AbdollahpouriBM19,DBLP:conf/recsys/AbdollahpouriBM17,DBLP:journals/pvldb/YinCLYC12,DBLP:conf/sigir/ZhuWC20,DBLP:conf/recsys/TsintzouPT19}, and \textit{Fairness}~\cite{deldjoo2020flexible,DBLP:conf/cikm/ZhuHC18}. It is worth mentioning that \elliot is the framework that exposes both the largest number of metrics and the only one considering bias and fairness measures. Moreover, the user can choose any metric to drive the model selection and the tuning. 
\textbf{\subsubsection{Statistical Tests}~\label{sec:framework_result_stats_test}}
\Cref{table:framework} shows that the reviewed related frameworks miss statistical hypothesis tests. This is probably due to the need to compute fine-grained (e.g., per-user or per-partition) results and retain them for each recommendation model. It implies that the framework should be designed for multi-recommender evaluation and handling the fine-grained results.
\elliot brings the opportunity to compute two statistical hypothesis tests, i.e., \textit{Wilcoxon} and \textit{Paired t-test}, activating a flag in the configuration file. 

\subsection{Framework Outcomes}~\label{sec:framework_output}
When the experiment finishes, it is time for \framework to collect the results through the \textit{Output} module in \Cref{fig:framework}.
\elliot gives the possibility to store three classes of output reports: (i) \textit{Performance Tables}, (ii) \textit{Model Weights}, and (iii) \textit{Recommendation Lists}. 
The former consist of spreadsheets (in a \textit{tab-separated-value} format) with all the metric values computed on the test set for every recommendation model specified in the configuration file. 
The tables comprise cut-off specific and model-specific tables (i.e., considering each combination of the explored parameters).
The user can also choose to store tables with the triple format, i.e., \textit{<Model, Metric, Value>}. 
Tables also include cut-off-specific statistical hypothesis tests and a JSON file that summarizes the best model parameters.
Optionally, \elliot saves model weights to avoid future re-training of the recommender.
Finally, \elliot stores the top-$k$ recommendation lists for each model adopting a tab-separated \textit{<User, Item, Predicted Score>} triple-based format.
\subsection{Preparation of the Experiment}~\label{sec:framework_config_file}
The operation of \elliot is triggered by a single configuration file written in YAML. 
Configuration~\ref{listing:helloworld} shows a toy example of a configuration file. 
The first section details the data loading, filtering, and splitting information as defined in~\Cref{sec:framework_data}. 
The \texttt{models} section represents the recommendation models configuration, e.g., Item-$k$NN, described in~\Cref{sec:framework_model_rec}.
Here, the model-specific hyperparameter optimization strategies are specified (\Cref{sec:framework_model_hyperopt}), e.g., the grid-search in Configuration~\ref{listing:helloworld}. 
The \texttt{evaluation} section details the evaluation strategy with the desired metrics (\Cref{sec:framework_result}), e.g., nDCG in the toy example. 
Finally, \texttt{save\_recs} and \texttt{top\_k} keys detail, for example, the \textit{Output} module abilities described in~\Cref{sec:framework_output}.
It is worth noticing that, to the best of our knowledge, \elliot is the only framework able to run an extensive set of reproducible experiments by merely preparing a single configuration file. 
\Cref{sec:experiments} exemplifies two real experimental scenarios commenting on the salient parts 
of the configuration files.

\definecolor{applegreen}{rgb}{0.55, 0.71, 0.0}
\begin{listing}[!t]
\caption{\small{\texttt{hello\_world.yml}}}
\begin{minted}[
    frame=lines,
    framerule=0.8pt,
    escapeinside=||,
    fontsize=\footnotesize,
  ]{yaml}
|\textbf{\textcolor{Black}{experiment}}|:
  |\textbf{\textcolor{Thistle}{dataset}}|: movielens_1m
  |\textbf{\textcolor{Thistle}{data\_config}}|:
    |\textbf{\textcolor{Thistle}{strategy}}|: dataset
    |\textbf{\textcolor{Thistle}{dataset\_path}}|: ../data/movielens_1m/dataset.tsv
  |\textbf{\textcolor{Thistle}{splitting}}|:
    |\textbf{\textcolor{Thistle}{test\_splitting}}|:
        |\textbf{\textcolor{Thistle}{strategy}}|: random_subsampling
        |\textbf{\textcolor{Thistle}{test\_ratio}}|: 0.2
  |\textbf{\textcolor{JungleGreen}{models}}|:
    |\textbf{\textcolor{JungleGreen}{ItemKNN}}|:
      |\textbf{\textcolor{JungleGreen}{meta}}|:
        |\textbf{\textcolor{JungleGreen}{hyper\_opt\_alg}}|: grid
        |\textbf{\textcolor{Orange}{save\_recs}}|: True
      |\textbf{\textcolor{JungleGreen}{neighbors}}|: |\textcolor{black}{[50, 100]}|
      |\textbf{\textcolor{JungleGreen}{similarity}}|: cosine
  |\textbf{\textcolor{applegreen}{evaluation}}|:
    |\textbf{\textcolor{applegreen}{simple\_metrics}}|: |\textcolor{black}{[nDCG]}|
  |\textbf{\textcolor{Orange}{top\_k}}|: 10
\end{minted}
\label{listing:helloworld}
\vspace{-1em}
\end{listing}

\section{Experimental Scenarios}~\label{sec:experiments}
We  illustrate how to prepare, execute and evaluate a \textit{basic} and a more \textit{advanced} experimental scenario with \elliot.

\subsection{Basic Configuration}\label{sec:standards}
\noindent \textbf{Experiment.} 
In the first scenario, the experiments require comparing a group of RSs whose parameters are optimized via a grid-search. 
Configuration \ref{lst:experiment1} specifies the data loading information, i.e., semantic features source files, in addition to the filtering and splitting strategies. 
In particular, the latter supplies an entirely automated way of preprocessing the dataset, which is often a time-consuming and non-easily-reproducible phase. 
The \mbox{\texttt{simple\_metrics}} field allows computing accuracy and beyond-accuracy metrics, with two top-$k$ cut-off values (5 and 10) by merely inserting the list of desired measures, e.g., \texttt{[Precision, nDCG, ...]}. 
The knowledge-aware recommendation model, \texttt{AttributeItemKNN}, is compared against two baselines: \texttt{Random} and \texttt{ItemKNN}, along with a user-implemented model that is \texttt{external.MostPop}. 
The configuration makes use of \elliot's feature of conducting a grid search-based hyperparameter optimization strategy by merely passing a list of possible hyperparameter values, e.g., \texttt{neighbors: [50, 70, 100]}. 
The reported models are selected according to \textit{nDCG@10}.

\noindent \textbf{Results.}
\Cref{table:exp2} displays a portion of experimental results generated by feeding \elliot with the configuration file. 
The table reports four metric values computed 
on recommendation lists at cutoffs $5$ and $10$
generated by the models selected after the hyperparameter tuning phase. 
For instance, Attribute-I-$k$NN model reports values for the configuration with \texttt{neighbors} set to \texttt{100} and \texttt{similarity} set to \texttt{braycurtis}. 
\Cref{table:exp2} confirms some common findings: the item coverage value ($ICov@10$) of an Attribute-I-$k$NN model is higher than the one measured on I-$k$NN, and I-$k$NN is the most accurate model.

\subsection{Advanced Configuration}

\noindent \textbf{Experiment.}
The second scenario depicts a more complex experimental setting. 
In Configuration~\ref{lst:experiment2}, the user specifies an elaborate data splitting strategy, i.e., \texttt{random\_subsampling} (for test splitting) and \texttt{random\_cross\_validation} (for model selection), by setting few splitting configuration fields. 
Configuration~\ref{lst:experiment2} does not provide a cut-off value, and thus a top-$k$ field value of 50 is assumed as the cut-off. 
Moreover, the evaluation section includes the \texttt{UserMADrating} metric.
\elliot considers it as a complex metric since it requires additional arguments (as shown in Configuration~\ref{lst:experiment2}).
The user also wants to implement a more advanced hyperparameter tuning optimization. For instance, regarding NeuMF, Bayesian optimization using \textit{Tree of Parzen Estimators}~\cite{DBLP:conf/nips/BergstraBBK11} is required (i.e., \texttt{\mbox{hyper\_opt\_alg}: tpe}) with a logarithmic uniform sampling for the learning rate search space.
Moreover, \elliot allows considering complex neural architecture search spaces by inserting lists of tuples. For instance, \texttt{(32, 16, 8)} indicates that the neural network consists of three hidden layers with $32$, $16$, and $8$ units, respectively.

\begin{listing}[!t]
\caption{\texttt{basic\_configuration.yml}}
\begin{minted}[
    frame=lines,
    framerule=0.8pt,
    escapeinside=||,
    fontsize=\footnotesize,
  ]{yaml}
|\textbf{\textcolor{black}{experiment}}|:
    |\textbf{\textcolor{Thistle}{dataset}}|: cat_dbpedia_movielens_1m
    |\textbf{\textcolor{Thistle}{data\_config}}|:
        |\textbf{\textcolor{Thistle}{strategy}}|: dataset
        |\textbf{\textcolor{Thistle}{dataloader}}|: KnowledgeChainsLoader
        |\textbf{\textcolor{Thistle}{dataset\_path}}|: <...>/dataset.tsv
        |\textbf{\textcolor{Thistle}{side\_information}}|:
            <...>
    |\textbf{\textcolor{Thistle}{prefiltering}}|:
        |\textbf{\textcolor{Thistle}{strategy}}|: user_average
    |\textbf{\textcolor{Thistle}{splitting}}|:
        |\textbf{\textcolor{Thistle}{test\_splitting}}|:
            |\textbf{\textcolor{Thistle}{strategy}}|: temporal_hold_out
            |\textbf{\textcolor{Thistle}{test\_ratio}}|: 0.2
        <...>
    |\textbf{\textcolor{JungleGreen}{external\_models\_path}}|: ../external/models/__init__.py
    |\textbf{\textcolor{JungleGreen}{models}}|:
        |\textbf{\textcolor{JungleGreen}{Random}}|:
            <...>
        |\textbf{\textcolor{JungleGreen}{external.MostPop}}|:
            <...>
        |\textbf{\textcolor{JungleGreen}{AttributeItemKNN}}|:
            |\textbf{\textcolor{JungleGreen}{neighbors}}|: |\textcolor{black}{[50, 70, 100]}|
            |\textbf{\textcolor{JungleGreen}{similarity}}|: |\textcolor{black}{[braycurtis, manhattan]}|
            <...>
    |\textbf{\textcolor{applegreen}{evaluation}}|:
        |\textbf{\textcolor{applegreen}{cutoffs}}|: |\textcolor{black}{[10, 5]}|
        |\textbf{\textcolor{applegreen}{evaluation}}|: |\textcolor{black}{[nDCG, Precision, ItemCoverage, EPC, Gini]}|
        |\textbf{\textcolor{applegreen}{relevance\_threshold}}|: 1
    |\textbf{\textcolor{Orange}{top\_k}}|: 50
\end{minted}
\vspace{-0.5em}
\begin{flushleft}
\scriptsize{\url{https://github.com/sisinflab/elliot/blob/master/config_files/basic_configuration.yml}}
\end{flushleft}
\hrule
\vspace{-1em}
\label{lst:experiment1}
\end{listing}
        

\begin{table}[t]
\caption{Experimental results for Configuration~\ref{lst:experiment1}.}
\label{table:exp2}
\small
{
\renewcommand{\arraystretch}{0.9}
\begin{tabularx}{\linewidth}{Xrrrr}
\toprule
\textbf{Model}  & \multicolumn{1}{l}{\textit{nDCG}@5} & \multicolumn{1}{l}{\textit{ICov}@5} & \multicolumn{1}{l}{\textit{nDCG}@10} & \multicolumn{1}{l}{\textit{ICov}@10} \\ 
\toprule
Random  & 0.0098 & 3197 & 0.0056 & 3197 \\ 
MostPop  & 0.0699 &	68 & 0.0728  & 96 \\ 
I-\textit{k}NN &  0.0791	& 448 & 0.0837 & 710 \\
Attribute-I-$k$NN  & 0.0464	&1575 & 0.0485  & 2102 \\ 
\bottomrule
\end{tabularx}
}
\vspace{-10pt}
\end{table}


\noindent \textbf{Results.}
\Cref{table:exp3} provides a summary of the experimental results obtained feeding \elliot with Configuration~\ref{lst:experiment2}. 
Even here, 
the columns report the values for all the considered metrics (simple and complex metrics). Configuration~\ref{lst:experiment2} also requires statistical hypothesis tests. Therefore, the table reports the \textit{Wilcoxon-test} outcome (computed on pairs of models with their best configuration). 
\iffalse
MultiVAE, coherently with the literature, outperforms the other baselines with the selected configuration (256 as batch size, intermediate and latent dimensions of $500$ and $100$ respectively, and learning rate set to $0.0005$).
\alejandro{these values (of the 'selected configuration') are confusing, as in the previous section, because they are not reported in the table, and in this case they are not even provided in the conf file. Should we mention them here or is it better to skip them? A third option would be to refer the reader to a github page where more details are provided...}~\felice{I would not report the values in the parenthesis also to gain a bit of space.}~\claudio{Yes, these values are no longer present in the listing 3}
\else
MultiVAE, coherently with the literature, outperforms the other baselines.
\fi


\section{Conclusion}~\label{sec:conclusion}

\begin{listing}[t]
\caption{\texttt{advanced\_configuration.yml}}
\begin{minted}[
    frame=lines,
    framerule=0.8pt,
    escapeinside=||,
    fontsize=\footnotesize,
  ]{yaml}
|\textbf{\textcolor{black}{experiment}}|:
    |\textbf{\textcolor{Thistle}{dataset}}|: movielens_1m
    |\textbf{\textcolor{Thistle}{data\_config}}|:
        |\textbf{\textcolor{Thistle}{strategy}}|: dataset
        |\textbf{\textcolor{Thistle}{dataset\_path}}|: <...>/dataset.tsv
    |\textbf{\textcolor{Thistle}{prefiltering}}|:
        |\textbf{\textcolor{Thistle}{strategy}}|: iterative_k_core
        |\textbf{\textcolor{Thistle}{core}}|: 10
    |\textbf{\textcolor{Thistle}{splitting}}|:
        |\textbf{\textcolor{Thistle}{test\_splitting}}|:
            |\textbf{\textcolor{Thistle}{strategy}}|: random_subsampling
            |\textbf{\textcolor{Thistle}{test\_ratio}}|: 0.2
        |\textbf{\textcolor{Thistle}{validation\_splitting}}|:
            |\textbf{\textcolor{Thistle}{strategy}}|: random_cross_validation
            |\textbf{\textcolor{Thistle}{folds}}|: 5
    |\textbf{\textcolor{JungleGreen}{models}}|:
        |\textbf{\textcolor{JungleGreen}{BPRMF}}|:
            <...>
        |\textbf{\textcolor{JungleGreen}{NeuMF}}|:
            |\textbf{\textcolor{JungleGreen}{meta}}|:
                |\textbf{\textcolor{JungleGreen}{hyper\_max\_evals}}|: 5
                |\textbf{\textcolor{JungleGreen}{hyper\_opt\_alg}}|: tpe
            |\textbf{\textcolor{JungleGreen}{lr}}|: |\textcolor{black}{[loguniform, -10, -1]}|
            |\textbf{\textcolor{JungleGreen}{mf\_factors}}|: |\textcolor{black}{[quniform, 8, 32, 1]}|
            |\textbf{\textcolor{JungleGreen}{mlp\_hidden\_size}}|: |\textcolor{black}{[(32, 16, 8), (64, 32, 16)]}|
            <...>
        |\textbf{\textcolor{JungleGreen}{MultiVAE}}|:
            <...>
    |\textbf{\textcolor{applegreen}{evaluation}}|:
        |\textbf{\textcolor{applegreen}{simple\_metrics}}|: |\textcolor{black}{[nDCG, ARP, ACLT]}|
        |\textbf{\textcolor{applegreen}{wilcoxon\_test}}|: True
        |\textbf{\textcolor{applegreen}{complex\_metrics}}|: 
        - |\textbf{\textcolor{applegreen}{metric}}|: UserMADrating
            |\textbf{\textcolor{applegreen}{clustering\_name}}|: Happiness
            |\textbf{\textcolor{applegreen}{clustering\_file}}|: <...>/u_happy.tsv
        |\textbf{\textcolor{applegreen}{relevance\_threshold}}|: 1
    |\textbf{\textcolor{Orange}{top\_k}}|: 50
\end{minted}
\vspace{-0.5em}
\begin{flushleft}
\scriptsize{\url{https://github.com/sisinflab/elliot/blob/master/config_files/advanced_configuration.yml}}
\end{flushleft}
\hrule
\label{lst:experiment2}
\vspace{-1em}
\end{listing}
 \framework is a framework that examines the recommendation process from an RS researcher's perspective.
It requires the user just to compile a flexible configuration file to conduct a rigorous and reproducible experimental evaluation.
The framework provides several loading, prefiltering, splitting, hyperparameter optimization strategies, recommendation models, and statistical hypothesis tests.
\framework reports can be directly analyzed and inserted into research papers.
We reviewed the RS evaluation literature, positioning \framework among the existing frameworks, and highlighting its advantages and limitations.
Next, we explored the framework architecture and how to build a working (and reproducible) experimental benchmark.
To the best of our knowledge, \framework is the first recommendation framework that provides a full multi-recommender experimental pipeline based on a simple configuration file.
We plan to extend soon \framework in various directions to include: sequential recommendation scenarios, adversarial attacks, reinforcement learning-based recommendation systems, differential privacy facilities, sampled evaluation, and distributed recommendation.

\begin{table}[t]
\caption{Experimental results for Configuration~\ref{lst:experiment2}.}
\label{table:exp3}
\small
{
\renewcommand{\arraystretch}{0.9}
\setlength{\tabcolsep}{0.29em}
\begin{tabularx}{\columnwidth}{Xrrrr}
\toprule
\textbf{Model} & \multicolumn{1}{l}{\textit{nDCG}@50\phantom{$\dagger$}} & \multicolumn{1}{l}{\textit{ARP}@50\phantom{$\dagger$}} & \multicolumn{1}{l}{\textit{ACLT}@50\phantom{$\dagger$}} & \multicolumn{1}{l}{\textit{UMAD}$_H$@50} \\ 
\toprule
BPRMF & 0.2390\phantom{$\dagger$} & 1096\phantom{$\dagger$} & 0.0420\phantom{$\dagger$} & 0.0516 \\ 
NeuMF & 0.2585\phantom{$\dagger$} & 919\phantom{$\dagger$} & 0.8616\phantom{$\dagger$} & 0.0032 \\ 
MultiVAE & 0.2922$\dagger$ & 755$\dagger$ & 3.2871$\dagger$ & 0.1588 \\ 
\midrule
\multicolumn{5}{l}{\textit{$\dagger\text{p-value} \leq 0.001$ using Wilcoxon-test}} \\
\bottomrule
\end{tabularx}
}
    \vspace{-15pt}
\end{table}

\iftrue
\begin{acks}
The authors acknowledge partial support of the projects: Servizi Locali 2.0, PON ARS01\_00876 Bio-D, PON ARS01\_00821 FLET4.0, PON ARS01\_00917 OK-INSAID, H2020 PASSPARTOUT,
PID2019-108965GB-I00
\end{acks}
\fi

\bibliographystyle{ACM-Reference-Format}
\bibliography{bibliography}


\begin{thebibliography}{109}


\ifx \showCODEN    \undefined \def \showCODEN     #1{\unskip}     \fi
\ifx \showDOI      \undefined \def \showDOI       #1{#1}\fi
\ifx \showISBNx    \undefined \def \showISBNx     #1{\unskip}     \fi
\ifx \showISBNxiii \undefined \def \showISBNxiii  #1{\unskip}     \fi
\ifx \showISSN     \undefined \def \showISSN      #1{\unskip}     \fi
\ifx \showLCCN     \undefined \def \showLCCN      #1{\unskip}     \fi
\ifx \shownote     \undefined \def \shownote      #1{#1}          \fi
\ifx \showarticletitle \undefined \def \showarticletitle #1{#1}   \fi
\ifx \showURL      \undefined \def \showURL       {\relax}        \fi
\providecommand\bibfield[2]{#2}
\providecommand\bibinfo[2]{#2}
\providecommand\natexlab[1]{#1}
\providecommand\showeprint[2][]{arXiv:#2}

\bibitem[\protect\citeauthoryear{Abdollahpouri}{Abdollahpouri}{2019}]%
        {DBLP:conf/aies/Abdollahpouri19}
\bibfield{author}{\bibinfo{person}{Himan Abdollahpouri}.}
  \bibinfo{year}{2019}\natexlab{}.
\newblock \showarticletitle{Popularity Bias in Ranking and Recommendation}. In
  \bibinfo{booktitle}{\emph{Proceedings of the 2019 {AAAI/ACM} Conference on
  AI, Ethics, and Society, {AIES} 2019, Honolulu, HI, USA, January 27-28,
  2019}}, \bibfield{editor}{\bibinfo{person}{Vincent Conitzer},
  \bibinfo{person}{Gillian~K. Hadfield}, {and} \bibinfo{person}{Shannon
  Vallor}} (Eds.). \bibinfo{publisher}{{ACM}}, \bibinfo{pages}{529--530}.
\newblock


\bibitem[\protect\citeauthoryear{Abdollahpouri, Burke, and
  Mobasher}{Abdollahpouri et~al\mbox{.}}{2017}]%
        {DBLP:conf/recsys/AbdollahpouriBM17}
\bibfield{author}{\bibinfo{person}{Himan Abdollahpouri}, \bibinfo{person}{Robin
  Burke}, {and} \bibinfo{person}{Bamshad Mobasher}.}
  \bibinfo{year}{2017}\natexlab{}.
\newblock \showarticletitle{Controlling Popularity Bias in Learning-to-Rank
  Recommendation}. In \bibinfo{booktitle}{\emph{Proceedings of the Eleventh
  {ACM} Conference on Recommender Systems, RecSys 2017, Como, Italy, August
  27-31, 2017}}, \bibfield{editor}{\bibinfo{person}{Paolo Cremonesi},
  \bibinfo{person}{Francesco Ricci}, \bibinfo{person}{Shlomo Berkovsky}, {and}
  \bibinfo{person}{Alexander Tuzhilin}} (Eds.). \bibinfo{publisher}{{ACM}},
  \bibinfo{pages}{42--46}.
\newblock


\bibitem[\protect\citeauthoryear{Abdollahpouri, Burke, and
  Mobasher}{Abdollahpouri et~al\mbox{.}}{2019}]%
        {DBLP:conf/flairs/AbdollahpouriBM19}
\bibfield{author}{\bibinfo{person}{Himan Abdollahpouri}, \bibinfo{person}{Robin
  Burke}, {and} \bibinfo{person}{Bamshad Mobasher}.}
  \bibinfo{year}{2019}\natexlab{}.
\newblock \showarticletitle{Managing Popularity Bias in Recommender Systems
  with Personalized Re-Ranking}. In \bibinfo{booktitle}{\emph{Proceedings of
  the Thirty-Second International Florida Artificial Intelligence Research
  Society Conference, Sarasota, Florida, USA, May 19-22 2019}},
  \bibfield{editor}{\bibinfo{person}{Roman Bart{\'{a}}k} {and}
  \bibinfo{person}{Keith~W. Brawner}} (Eds.). \bibinfo{publisher}{{AAAI}
  Press}, \bibinfo{pages}{413--418}.
\newblock


\bibitem[\protect\citeauthoryear{Anelli, Delic, Sottocornola, Smith, Andrade,
  Belli, Bronstein, Gupta, Ktena, Lung{-}Yut{-}Fong, Portman, Tejani, Xie, Zhu,
  and Shi}{Anelli et~al\mbox{.}}{2020}]%
        {DBLP:conf/recsys/AnelliDSSABBGKL20}
\bibfield{author}{\bibinfo{person}{Vito~Walter Anelli}, \bibinfo{person}{Amra
  Delic}, \bibinfo{person}{Gabriele Sottocornola}, \bibinfo{person}{Jessie
  Smith}, \bibinfo{person}{Nazareno Andrade}, \bibinfo{person}{Luca Belli},
  \bibinfo{person}{Michael~M. Bronstein}, \bibinfo{person}{Akshay Gupta},
  \bibinfo{person}{Sofia~Ira Ktena}, \bibinfo{person}{Alexandre
  Lung{-}Yut{-}Fong}, \bibinfo{person}{Frank Portman}, \bibinfo{person}{Alykhan
  Tejani}, \bibinfo{person}{Yuanpu Xie}, \bibinfo{person}{Xiao Zhu}, {and}
  \bibinfo{person}{Wenzhe Shi}.} \bibinfo{year}{2020}\natexlab{}.
\newblock \showarticletitle{RecSys 2020 Challenge Workshop: Engagement
  Prediction on Twitter's Home Timeline}. In \bibinfo{booktitle}{\emph{RecSys
  2020: Fourteenth {ACM} Conference on Recommender Systems, Virtual Event,
  Brazil, September 22-26, 2020}}, \bibfield{editor}{\bibinfo{person}{Rodrygo
  L.~T. Santos}, \bibinfo{person}{Leandro~Balby Marinho},
  \bibinfo{person}{Elizabeth~M. Daly}, \bibinfo{person}{Li~Chen},
  \bibinfo{person}{Kim Falk}, \bibinfo{person}{Noam Koenigstein}, {and}
  \bibinfo{person}{Edleno~Silva de~Moura}} (Eds.). \bibinfo{publisher}{{ACM}},
  \bibinfo{pages}{623--627}.
\newblock


\bibitem[\protect\citeauthoryear{{Anelli}, {Di Noia}, {Di Sciascio}, {Ragone},
  and {Trotta}}{{Anelli} et~al\mbox{.}}{2020}]%
        {9143460}
\bibfield{author}{\bibinfo{person}{V.~W. {Anelli}}, \bibinfo{person}{T. {Di
  Noia}}, \bibinfo{person}{E. {Di Sciascio}}, \bibinfo{person}{A. {Ragone}},
  {and} \bibinfo{person}{J. {Trotta}}.} \bibinfo{year}{2020}\natexlab{}.
\newblock \showarticletitle{Semantic Interpretation of Top-N Recommendations}.
\newblock \bibinfo{journal}{\emph{IEEE Transactions on Knowledge and Data
  Engineering}} (\bibinfo{year}{2020}), \bibinfo{pages}{1--1}.
\newblock


\bibitem[\protect\citeauthoryear{Anelli, Noia, Sciascio, Pomo, and
  Ragone}{Anelli et~al\mbox{.}}{2019a}]%
        {DBLP:conf/recsys/AnelliNSPR19}
\bibfield{author}{\bibinfo{person}{Vito~Walter Anelli},
  \bibinfo{person}{Tommaso~Di Noia}, \bibinfo{person}{Eugenio~Di Sciascio},
  \bibinfo{person}{Claudio Pomo}, {and} \bibinfo{person}{Azzurra Ragone}.}
  \bibinfo{year}{2019}\natexlab{a}.
\newblock \showarticletitle{On the discriminative power of hyper-parameters in
  cross-validation and how to choose them}. In
  \bibinfo{booktitle}{\emph{Proceedings of the 13th {ACM} Conference on
  Recommender Systems, RecSys 2019, Copenhagen, Denmark, September 16-20,
  2019}}, \bibfield{editor}{\bibinfo{person}{Toine Bogers},
  \bibinfo{person}{Alan Said}, \bibinfo{person}{Peter Brusilovsky}, {and}
  \bibinfo{person}{Domonkos Tikk}} (Eds.). \bibinfo{publisher}{{ACM}},
  \bibinfo{pages}{447--451}.
\newblock


\bibitem[\protect\citeauthoryear{Anelli, Noia, Sciascio, Ragone, and
  Trotta}{Anelli et~al\mbox{.}}{2019b}]%
        {DBLP:conf/semweb/AnelliNSRT19}
\bibfield{author}{\bibinfo{person}{Vito~Walter Anelli},
  \bibinfo{person}{Tommaso~Di Noia}, \bibinfo{person}{Eugenio~Di Sciascio},
  \bibinfo{person}{Azzurra Ragone}, {and} \bibinfo{person}{Joseph Trotta}.}
  \bibinfo{year}{2019}\natexlab{b}.
\newblock \showarticletitle{How to Make Latent Factors Interpretable by Feeding
  Factorization Machines with Knowledge Graphs}. In
  \bibinfo{booktitle}{\emph{The Semantic Web - {ISWC} 2019 - 18th International
  Semantic Web Conference, Auckland, New Zealand, October 26-30, 2019,
  Proceedings, Part {I}}} \emph{(\bibinfo{series}{Lecture Notes in Computer
  Science}, Vol.~\bibinfo{volume}{11778})},
  \bibfield{editor}{\bibinfo{person}{Chiara Ghidini}, \bibinfo{person}{Olaf
  Hartig}, \bibinfo{person}{Maria Maleshkova}, \bibinfo{person}{Vojtech
  Sv{\'{a}}tek}, \bibinfo{person}{Isabel~F. Cruz}, \bibinfo{person}{Aidan
  Hogan}, \bibinfo{person}{Jie Song}, \bibinfo{person}{Maxime
  Lefran{\c{c}}ois}, {and} \bibinfo{person}{Fabien Gandon}} (Eds.).
  \bibinfo{publisher}{Springer}, \bibinfo{pages}{38--56}.
\newblock


\bibitem[\protect\citeauthoryear{Anelli, Noia, Sciascio, Ragone, and
  Trotta}{Anelli et~al\mbox{.}}{2019c}]%
        {DBLP:conf/ecir/AnelliNSRT19}
\bibfield{author}{\bibinfo{person}{Vito~Walter Anelli},
  \bibinfo{person}{Tommaso~Di Noia}, \bibinfo{person}{Eugenio~Di Sciascio},
  \bibinfo{person}{Azzurra Ragone}, {and} \bibinfo{person}{Joseph Trotta}.}
  \bibinfo{year}{2019}\natexlab{c}.
\newblock \showarticletitle{Local Popularity and Time in top-N Recommendation}.
  In \bibinfo{booktitle}{\emph{Advances in Information Retrieval - 41st
  European Conference on {IR} Research, {ECIR} 2019, Cologne, Germany, April
  14-18, 2019, Proceedings, Part {I}}} \emph{(\bibinfo{series}{Lecture Notes in
  Computer Science}, Vol.~\bibinfo{volume}{11437})},
  \bibfield{editor}{\bibinfo{person}{Leif Azzopardi}, \bibinfo{person}{Benno
  Stein}, \bibinfo{person}{Norbert Fuhr}, \bibinfo{person}{Philipp Mayr},
  \bibinfo{person}{Claudia Hauff}, {and} \bibinfo{person}{Djoerd Hiemstra}}
  (Eds.). \bibinfo{publisher}{Springer}, \bibinfo{pages}{861--868}.
\newblock


\bibitem[\protect\citeauthoryear{Baeza{-}Yates}{Baeza{-}Yates}{2020}]%
        {DBLP:conf/recsys/Baeza-Yates20}
\bibfield{author}{\bibinfo{person}{Ricardo Baeza{-}Yates}.}
  \bibinfo{year}{2020}\natexlab{}.
\newblock \showarticletitle{Bias in Search and Recommender Systems}. In
  \bibinfo{booktitle}{\emph{RecSys 2020: Fourteenth {ACM} Conference on
  Recommender Systems, Virtual Event, Brazil, September 22-26, 2020}},
  \bibfield{editor}{\bibinfo{person}{Rodrygo L.~T. Santos},
  \bibinfo{person}{Leandro~Balby Marinho}, \bibinfo{person}{Elizabeth~M. Daly},
  \bibinfo{person}{Li~Chen}, \bibinfo{person}{Kim Falk}, \bibinfo{person}{Noam
  Koenigstein}, {and} \bibinfo{person}{Edleno~Silva de~Moura}} (Eds.).
  \bibinfo{publisher}{{ACM}}, \bibinfo{pages}{2}.
\newblock


\bibitem[\protect\citeauthoryear{Bellog{\'{\i}}n, Castells, and
  Cantador}{Bellog{\'{\i}}n et~al\mbox{.}}{2011}]%
        {DBLP:conf/recsys/BelloginCC11}
\bibfield{author}{\bibinfo{person}{Alejandro Bellog{\'{\i}}n},
  \bibinfo{person}{Pablo Castells}, {and} \bibinfo{person}{Iv{\'{a}}n
  Cantador}.} \bibinfo{year}{2011}\natexlab{}.
\newblock \showarticletitle{Precision-oriented evaluation of recommender
  systems: an algorithmic comparison}. In \bibinfo{booktitle}{\emph{Proceedings
  of the 2011 {ACM} Conference on Recommender Systems, RecSys 2011, Chicago,
  IL, USA, October 23-27, 2011}}, \bibfield{editor}{\bibinfo{person}{Bamshad
  Mobasher}, \bibinfo{person}{Robin~D. Burke}, \bibinfo{person}{Dietmar
  Jannach}, {and} \bibinfo{person}{Gediminas Adomavicius}} (Eds.).
  \bibinfo{publisher}{{ACM}}, \bibinfo{pages}{333--336}.
\newblock


\bibitem[\protect\citeauthoryear{Bellog{\'{\i}}n and Said}{Bellog{\'{\i}}n and
  Said}{2021}]%
        {DBLP:journals/corr/abs-2102-00482}
\bibfield{author}{\bibinfo{person}{Alejandro Bellog{\'{\i}}n} {and}
  \bibinfo{person}{Alan Said}.} \bibinfo{year}{2021}\natexlab{}.
\newblock \showarticletitle{Improving Accountability in Recommender Systems
  Research Through Reproducibility}.
\newblock \bibinfo{journal}{\emph{CoRR}}  \bibinfo{volume}{abs/2102.00482}
  (\bibinfo{year}{2021}).
\newblock


\bibitem[\protect\citeauthoryear{Bellog{\'{\i}}n and
  S{\'{a}}nchez}{Bellog{\'{\i}}n and S{\'{a}}nchez}{2017}]%
        {DBLP:conf/recsys/BelloginS17}
\bibfield{author}{\bibinfo{person}{Alejandro Bellog{\'{\i}}n} {and}
  \bibinfo{person}{Pablo S{\'{a}}nchez}.} \bibinfo{year}{2017}\natexlab{}.
\newblock \showarticletitle{Revisiting Neighbourhood-Based Recommenders For
  Temporal Scenarios}. In \bibinfo{booktitle}{\emph{Proceedings of the 1st
  Workshop on Temporal Reasoning in Recommender Systems co-located with 11th
  International Conference on Recommender Systems (RecSys 2017), Como, Italy,
  August 27-31, 2017}} \emph{(\bibinfo{series}{{CEUR} Workshop Proceedings},
  Vol.~\bibinfo{volume}{1922})}, \bibfield{editor}{\bibinfo{person}{M{\'{a}}ria
  Bielikov{\'{a}}}, \bibinfo{person}{Veronika Bogina}, \bibinfo{person}{Tsvi
  Kuflik}, {and} \bibinfo{person}{Roy Sasson}} (Eds.).
  \bibinfo{publisher}{CEUR-WS.org}, \bibinfo{pages}{40--44}.
\newblock


\bibitem[\protect\citeauthoryear{Bennett and Lanning}{Bennett and
  Lanning}{2007}]%
        {bennett2007netflix}
\bibfield{author}{\bibinfo{person}{James Bennett} {and} \bibinfo{person}{Stan
  Lanning}.} \bibinfo{year}{2007}\natexlab{}.
\newblock \showarticletitle{The netflix prize}. In
  \bibinfo{booktitle}{\emph{Proceedings of the 13th {ACM} {SIGKDD}
  International Conference on Knowledge Discovery and Data Mining, San Jose,
  California, USA, August 12-15, 2007}}. \bibinfo{publisher}{{ACM}}.
\newblock


\bibitem[\protect\citeauthoryear{Bergstra, Bardenet, Bengio, and
  K{\'{e}}gl}{Bergstra et~al\mbox{.}}{2011}]%
        {DBLP:conf/nips/BergstraBBK11}
\bibfield{author}{\bibinfo{person}{James Bergstra}, \bibinfo{person}{R{\'{e}}mi
  Bardenet}, \bibinfo{person}{Yoshua Bengio}, {and}
  \bibinfo{person}{Bal{\'{a}}zs K{\'{e}}gl}.} \bibinfo{year}{2011}\natexlab{}.
\newblock \showarticletitle{Algorithms for Hyper-Parameter Optimization}. In
  \bibinfo{booktitle}{\emph{Advances in Neural Information Processing Systems
  24: 25th Annual Conference on Neural Information Processing Systems 2011.
  Proceedings of a meeting held 12-14 December 2011, Granada, Spain}},
  \bibfield{editor}{\bibinfo{person}{John Shawe{-}Taylor},
  \bibinfo{person}{Richard~S. Zemel}, \bibinfo{person}{Peter~L. Bartlett},
  \bibinfo{person}{Fernando C.~N. Pereira}, {and} \bibinfo{person}{Kilian~Q.
  Weinberger}} (Eds.). \bibinfo{pages}{2546--2554}.
\newblock


\bibitem[\protect\citeauthoryear{Bergstra, Yamins, and Cox}{Bergstra
  et~al\mbox{.}}{2013}]%
        {DBLP:conf/icml/BergstraYC13}
\bibfield{author}{\bibinfo{person}{James Bergstra}, \bibinfo{person}{Daniel
  Yamins}, {and} \bibinfo{person}{David~D. Cox}.}
  \bibinfo{year}{2013}\natexlab{}.
\newblock \showarticletitle{Making a Science of Model Search: Hyperparameter
  Optimization in Hundreds of Dimensions for Vision Architectures}. In
  \bibinfo{booktitle}{\emph{Proceedings of the 30th International Conference on
  Machine Learning, {ICML} 2013, Atlanta, GA, USA, 16-21 June 2013}}
  \emph{(\bibinfo{series}{{JMLR} Workshop and Conference Proceedings},
  Vol.~\bibinfo{volume}{28})}. \bibinfo{publisher}{JMLR.org},
  \bibinfo{pages}{115--123}.
\newblock


\bibitem[\protect\citeauthoryear{Campos, D{\'{\i}}ez, and Cantador}{Campos
  et~al\mbox{.}}{2014}]%
        {DBLP:journals/umuai/CamposDC14}
\bibfield{author}{\bibinfo{person}{Pedro~G. Campos}, \bibinfo{person}{Fernando
  D{\'{\i}}ez}, {and} \bibinfo{person}{Iv{\'{a}}n Cantador}.}
  \bibinfo{year}{2014}\natexlab{}.
\newblock \showarticletitle{Time-aware recommender systems: a comprehensive
  survey and analysis of existing evaluation protocols}.
\newblock \bibinfo{journal}{\emph{User Model. User Adapt. Interact.}}
  \bibinfo{volume}{24}, \bibinfo{number}{1-2} (\bibinfo{year}{2014}),
  \bibinfo{pages}{67--119}.
\newblock


\bibitem[\protect\citeauthoryear{Castells, Hurley, and Vargas}{Castells
  et~al\mbox{.}}{2015}]%
        {DBLP:reference/sp/CastellsHV15}
\bibfield{author}{\bibinfo{person}{Pablo Castells}, \bibinfo{person}{Neil~J.
  Hurley}, {and} \bibinfo{person}{Saul Vargas}.}
  \bibinfo{year}{2015}\natexlab{}.
\newblock \showarticletitle{Novelty and Diversity in Recommender Systems}.
\newblock In \bibinfo{booktitle}{\emph{Recommender Systems Handbook}},
  \bibfield{editor}{\bibinfo{person}{Francesco Ricci}, \bibinfo{person}{Lior
  Rokach}, {and} \bibinfo{person}{Bracha Shapira}} (Eds.).
  \bibinfo{publisher}{Springer}, \bibinfo{pages}{881--918}.
\newblock


\bibitem[\protect\citeauthoryear{Chae, Kang, Kim, and Lee}{Chae
  et~al\mbox{.}}{2018}]%
        {DBLP:conf/cikm/ChaeKKL18}
\bibfield{author}{\bibinfo{person}{Dong{-}Kyu Chae}, \bibinfo{person}{Jin{-}Soo
  Kang}, \bibinfo{person}{Sang{-}Wook Kim}, {and} \bibinfo{person}{Jung{-}Tae
  Lee}.} \bibinfo{year}{2018}\natexlab{}.
\newblock \showarticletitle{{CFGAN:} {A} Generic Collaborative Filtering
  Framework based on Generative Adversarial Networks}. In
  \bibinfo{booktitle}{\emph{Proceedings of the 27th {ACM} International
  Conference on Information and Knowledge Management, {CIKM} 2018, Torino,
  Italy, October 22-26, 2018}}, \bibfield{editor}{\bibinfo{person}{Alfredo
  Cuzzocrea}, \bibinfo{person}{James Allan}, \bibinfo{person}{Norman~W. Paton},
  \bibinfo{person}{Divesh Srivastava}, \bibinfo{person}{Rakesh Agrawal},
  \bibinfo{person}{Andrei~Z. Broder}, \bibinfo{person}{Mohammed~J. Zaki},
  \bibinfo{person}{K.~Sel{\c{c}}uk Candan}, \bibinfo{person}{Alexandros
  Labrinidis}, \bibinfo{person}{Assaf Schuster}, {and} \bibinfo{person}{Haixun
  Wang}} (Eds.). \bibinfo{publisher}{{ACM}}, \bibinfo{pages}{137--146}.
\newblock


\bibitem[\protect\citeauthoryear{Chen, Zhang, He, Nie, Liu, and Chua}{Chen
  et~al\mbox{.}}{2017}]%
        {DBLP:conf/sigir/ChenZ0NLC17}
\bibfield{author}{\bibinfo{person}{Jingyuan Chen}, \bibinfo{person}{Hanwang
  Zhang}, \bibinfo{person}{Xiangnan He}, \bibinfo{person}{Liqiang Nie},
  \bibinfo{person}{Wei Liu}, {and} \bibinfo{person}{Tat{-}Seng Chua}.}
  \bibinfo{year}{2017}\natexlab{}.
\newblock \showarticletitle{Attentive Collaborative Filtering: Multimedia
  Recommendation with Item- and Component-Level Attention}. In
  \bibinfo{booktitle}{\emph{Proceedings of the 40th International {ACM} {SIGIR}
  Conference on Research and Development in Information Retrieval, Shinjuku,
  Tokyo, Japan, August 7-11, 2017}}, \bibfield{editor}{\bibinfo{person}{Noriko
  Kando}, \bibinfo{person}{Tetsuya Sakai}, \bibinfo{person}{Hideo Joho},
  \bibinfo{person}{Hang Li}, \bibinfo{person}{Arjen~P. de~Vries}, {and}
  \bibinfo{person}{Ryen~W. White}} (Eds.). \bibinfo{publisher}{{ACM}},
  \bibinfo{pages}{335--344}.
\newblock


\bibitem[\protect\citeauthoryear{Cheng, Koc, Harmsen, Shaked, Chandra, Aradhye,
  Anderson, Corrado, Chai, Ispir, Anil, Haque, Hong, Jain, Liu, and Shah}{Cheng
  et~al\mbox{.}}{2016}]%
        {DBLP:conf/recsys/Cheng0HSCAACCIA16}
\bibfield{author}{\bibinfo{person}{Heng{-}Tze Cheng}, \bibinfo{person}{Levent
  Koc}, \bibinfo{person}{Jeremiah Harmsen}, \bibinfo{person}{Tal Shaked},
  \bibinfo{person}{Tushar Chandra}, \bibinfo{person}{Hrishi Aradhye},
  \bibinfo{person}{Glen Anderson}, \bibinfo{person}{Greg Corrado},
  \bibinfo{person}{Wei Chai}, \bibinfo{person}{Mustafa Ispir},
  \bibinfo{person}{Rohan Anil}, \bibinfo{person}{Zakaria Haque},
  \bibinfo{person}{Lichan Hong}, \bibinfo{person}{Vihan Jain},
  \bibinfo{person}{Xiaobing Liu}, {and} \bibinfo{person}{Hemal Shah}.}
  \bibinfo{year}{2016}\natexlab{}.
\newblock \showarticletitle{Wide {\&} Deep Learning for Recommender Systems}.
  In \bibinfo{booktitle}{\emph{Proceedings of the 1st Workshop on Deep Learning
  for Recommender Systems, DLRS@RecSys 2016, Boston, MA, USA, September 15,
  2016}}, \bibfield{editor}{\bibinfo{person}{Alexandros Karatzoglou},
  \bibinfo{person}{Bal{\'{a}}zs Hidasi}, \bibinfo{person}{Domonkos Tikk},
  \bibinfo{person}{Oren~Sar Shalom}, \bibinfo{person}{Haggai Roitman},
  \bibinfo{person}{Bracha Shapira}, {and} \bibinfo{person}{Lior Rokach}}
  (Eds.). \bibinfo{publisher}{{ACM}}, \bibinfo{pages}{7--10}.
\newblock


\bibitem[\protect\citeauthoryear{Cremonesi, Koren, and Turrin}{Cremonesi
  et~al\mbox{.}}{2010}]%
        {DBLP:conf/recsys/CremonesiKT10}
\bibfield{author}{\bibinfo{person}{Paolo Cremonesi}, \bibinfo{person}{Yehuda
  Koren}, {and} \bibinfo{person}{Roberto Turrin}.}
  \bibinfo{year}{2010}\natexlab{}.
\newblock \showarticletitle{Performance of recommender algorithms on top-n
  recommendation tasks}. In \bibinfo{booktitle}{\emph{Proceedings of the 2010
  {ACM} Conference on Recommender Systems, RecSys 2010, Barcelona, Spain,
  September 26-30, 2010}}, \bibfield{editor}{\bibinfo{person}{Xavier
  Amatriain}, \bibinfo{person}{Marc Torrens}, \bibinfo{person}{Paul Resnick},
  {and} \bibinfo{person}{Markus Zanker}} (Eds.). \bibinfo{publisher}{{ACM}},
  \bibinfo{pages}{39--46}.
\newblock


\bibitem[\protect\citeauthoryear{Dacrema, Cremonesi, and Jannach}{Dacrema
  et~al\mbox{.}}{2019}]%
        {DBLP:conf/recsys/DacremaCJ19}
\bibfield{author}{\bibinfo{person}{Maurizio~Ferrari Dacrema},
  \bibinfo{person}{Paolo Cremonesi}, {and} \bibinfo{person}{Dietmar Jannach}.}
  \bibinfo{year}{2019}\natexlab{}.
\newblock \showarticletitle{Are we really making much progress? {A} worrying
  analysis of recent neural recommendation approaches}. In
  \bibinfo{booktitle}{\emph{Proceedings of the 13th {ACM} Conference on
  Recommender Systems, RecSys 2019, Copenhagen, Denmark, September 16-20,
  2019}}, \bibfield{editor}{\bibinfo{person}{Toine Bogers},
  \bibinfo{person}{Alan Said}, \bibinfo{person}{Peter Brusilovsky}, {and}
  \bibinfo{person}{Domonkos Tikk}} (Eds.). \bibinfo{publisher}{{ACM}},
  \bibinfo{pages}{101--109}.
\newblock


\bibitem[\protect\citeauthoryear{Deldjoo, Anelli, Zamani, Bellogin, and
  Di~Noia}{Deldjoo et~al\mbox{.}}{2020}]%
        {deldjoo2020flexible}
\bibfield{author}{\bibinfo{person}{Yashar Deldjoo},
  \bibinfo{person}{Vito~Walter Anelli}, \bibinfo{person}{Hamed Zamani},
  \bibinfo{person}{Alejandro Bellogin}, {and} \bibinfo{person}{Tommaso
  Di~Noia}.} \bibinfo{year}{2020}\natexlab{}.
\newblock \showarticletitle{A flexible framework for evaluating user and item
  fairness in recommender systems}.
\newblock \bibinfo{journal}{\emph{User Modeling and User-Adapted Interaction}}
  (\bibinfo{year}{2020}), \bibinfo{pages}{1--47}.
\newblock


\bibitem[\protect\citeauthoryear{Ekstrand}{Ekstrand}{2020}]%
        {DBLP:conf/cikm/Ekstrand20}
\bibfield{author}{\bibinfo{person}{Michael~D. Ekstrand}.}
  \bibinfo{year}{2020}\natexlab{}.
\newblock \showarticletitle{LensKit for Python: Next-Generation Software for
  Recommender Systems Experiments}. In \bibinfo{booktitle}{\emph{{CIKM} '20:
  The 29th {ACM} International Conference on Information and Knowledge
  Management, Virtual Event, Ireland, October 19-23, 2020}},
  \bibfield{editor}{\bibinfo{person}{Mathieu d'Aquin}, \bibinfo{person}{Stefan
  Dietze}, \bibinfo{person}{Claudia Hauff}, \bibinfo{person}{Edward Curry},
  {and} \bibinfo{person}{Philippe Cudr{\'{e}}{-}Mauroux}} (Eds.).
  \bibinfo{publisher}{{ACM}}, \bibinfo{pages}{2999--3006}.
\newblock


\bibitem[\protect\citeauthoryear{Ekstrand, Burke, and Diaz}{Ekstrand
  et~al\mbox{.}}{2019a}]%
        {DBLP:conf/recsys/EkstrandBD19}
\bibfield{author}{\bibinfo{person}{Michael~D. Ekstrand}, \bibinfo{person}{Robin
  Burke}, {and} \bibinfo{person}{Fernando Diaz}.}
  \bibinfo{year}{2019}\natexlab{a}.
\newblock \showarticletitle{Fairness and discrimination in recommendation and
  retrieval}. In \bibinfo{booktitle}{\emph{Proceedings of the 13th {ACM}
  Conference on Recommender Systems, RecSys 2019, Copenhagen, Denmark,
  September 16-20, 2019}}, \bibfield{editor}{\bibinfo{person}{Toine Bogers},
  \bibinfo{person}{Alan Said}, \bibinfo{person}{Peter Brusilovsky}, {and}
  \bibinfo{person}{Domonkos Tikk}} (Eds.). \bibinfo{publisher}{{ACM}},
  \bibinfo{pages}{576--577}.
\newblock


\bibitem[\protect\citeauthoryear{Ekstrand, Burke, and Diaz}{Ekstrand
  et~al\mbox{.}}{2019b}]%
        {DBLP:conf/sigir/EkstrandBD19}
\bibfield{author}{\bibinfo{person}{Michael~D. Ekstrand}, \bibinfo{person}{Robin
  Burke}, {and} \bibinfo{person}{Fernando Diaz}.}
  \bibinfo{year}{2019}\natexlab{b}.
\newblock \showarticletitle{Fairness and Discrimination in Retrieval and
  Recommendation}. In \bibinfo{booktitle}{\emph{Proceedings of the 42nd
  International {ACM} {SIGIR} Conference on Research and Development in
  Information Retrieval, {SIGIR} 2019, Paris, France, July 21-25, 2019}},
  \bibfield{editor}{\bibinfo{person}{Benjamin Piwowarski}, \bibinfo{person}{Max
  Chevalier}, \bibinfo{person}{{\'{E}}ric Gaussier}, \bibinfo{person}{Yoelle
  Maarek}, \bibinfo{person}{Jian{-}Yun Nie}, {and} \bibinfo{person}{Falk
  Scholer}} (Eds.). \bibinfo{publisher}{{ACM}}, \bibinfo{pages}{1403--1404}.
\newblock


\bibitem[\protect\citeauthoryear{Ekstrand, Ludwig, Konstan, and Riedl}{Ekstrand
  et~al\mbox{.}}{2011}]%
        {DBLP:conf/recsys/EkstrandLKR11}
\bibfield{author}{\bibinfo{person}{Michael~D. Ekstrand},
  \bibinfo{person}{Michael Ludwig}, \bibinfo{person}{Joseph~A. Konstan}, {and}
  \bibinfo{person}{John Riedl}.} \bibinfo{year}{2011}\natexlab{}.
\newblock \showarticletitle{Rethinking the recommender research ecosystem:
  reproducibility, openness, and LensKit}. In
  \bibinfo{booktitle}{\emph{Proceedings of the 2011 {ACM} Conference on
  Recommender Systems, RecSys 2011, Chicago, IL, USA, October 23-27, 2011}},
  \bibfield{editor}{\bibinfo{person}{Bamshad Mobasher},
  \bibinfo{person}{Robin~D. Burke}, \bibinfo{person}{Dietmar Jannach}, {and}
  \bibinfo{person}{Gediminas Adomavicius}} (Eds.). \bibinfo{publisher}{{ACM}},
  \bibinfo{pages}{133--140}.
\newblock


\bibitem[\protect\citeauthoryear{Frederickson}{Frederickson}{2018}]%
        {frederickson2018fast}
\bibfield{author}{\bibinfo{person}{Ben Frederickson}.}
  \bibinfo{year}{2018}\natexlab{}.
\newblock \bibinfo{title}{Fast python collaborative filtering for implicit
  datasets}.
\newblock
\newblock


\bibitem[\protect\citeauthoryear{Funk}{Funk}{2006}]%
        {funk2006netflix}
\bibfield{author}{\bibinfo{person}{Simon Funk}.}
  \bibinfo{year}{2006}\natexlab{}.
\newblock \bibinfo{title}{Netflix update: Try this at home}.
\newblock
\newblock


\bibitem[\protect\citeauthoryear{Gantner, Drumond, Freudenthaler, and
  Schmidt{-}Thieme}{Gantner et~al\mbox{.}}{2012}]%
        {DBLP:journals/jmlr/GantnerDFS12}
\bibfield{author}{\bibinfo{person}{Zeno Gantner}, \bibinfo{person}{Lucas
  Drumond}, \bibinfo{person}{Christoph Freudenthaler}, {and}
  \bibinfo{person}{Lars Schmidt{-}Thieme}.} \bibinfo{year}{2012}\natexlab{}.
\newblock \showarticletitle{Personalized Ranking for Non-Uniformly Sampled
  Items}. In \bibinfo{booktitle}{\emph{Proceedings of {KDD} Cup 2011
  competition, San Diego, CA, USA, 2011}} \emph{(\bibinfo{series}{{JMLR}
  Proceedings}, Vol.~\bibinfo{volume}{18})},
  \bibfield{editor}{\bibinfo{person}{Gideon Dror}, \bibinfo{person}{Yehuda
  Koren}, {and} \bibinfo{person}{Markus Weimer}} (Eds.).
  \bibinfo{publisher}{JMLR.org}, \bibinfo{pages}{231--247}.
\newblock


\bibitem[\protect\citeauthoryear{Gantner, Rendle, Freudenthaler, and
  Schmidt{-}Thieme}{Gantner et~al\mbox{.}}{2011}]%
        {DBLP:conf/recsys/GantnerRFS11}
\bibfield{author}{\bibinfo{person}{Zeno Gantner}, \bibinfo{person}{Steffen
  Rendle}, \bibinfo{person}{Christoph Freudenthaler}, {and}
  \bibinfo{person}{Lars Schmidt{-}Thieme}.} \bibinfo{year}{2011}\natexlab{}.
\newblock \showarticletitle{MyMediaLite: a free recommender system library}. In
  \bibinfo{booktitle}{\emph{Proceedings of the 2011 {ACM} Conference on
  Recommender Systems, RecSys 2011, Chicago, IL, USA, October 23-27, 2011}},
  \bibfield{editor}{\bibinfo{person}{Bamshad Mobasher},
  \bibinfo{person}{Robin~D. Burke}, \bibinfo{person}{Dietmar Jannach}, {and}
  \bibinfo{person}{Gediminas Adomavicius}} (Eds.). \bibinfo{publisher}{{ACM}},
  \bibinfo{pages}{305--308}.
\newblock


\bibitem[\protect\citeauthoryear{Gunawardana and Shani}{Gunawardana and
  Shani}{2015}]%
        {DBLP:reference/sp/GunawardanaS15}
\bibfield{author}{\bibinfo{person}{Asela Gunawardana} {and}
  \bibinfo{person}{Guy Shani}.} \bibinfo{year}{2015}\natexlab{}.
\newblock \showarticletitle{Evaluating Recommender Systems}.
\newblock In \bibinfo{booktitle}{\emph{Recommender Systems Handbook}},
  \bibfield{editor}{\bibinfo{person}{Francesco Ricci}, \bibinfo{person}{Lior
  Rokach}, {and} \bibinfo{person}{Bracha Shapira}} (Eds.).
  \bibinfo{publisher}{Springer}, \bibinfo{pages}{265--308}.
\newblock


\bibitem[\protect\citeauthoryear{Guo, Zhang, Sun, and Yorke{-}Smith}{Guo
  et~al\mbox{.}}{2015}]%
        {DBLP:conf/um/GuoZSY15}
\bibfield{author}{\bibinfo{person}{Guibing Guo}, \bibinfo{person}{Jie Zhang},
  \bibinfo{person}{Zhu Sun}, {and} \bibinfo{person}{Neil Yorke{-}Smith}.}
  \bibinfo{year}{2015}\natexlab{}.
\newblock \showarticletitle{LibRec: {A} Java Library for Recommender Systems}.
  In \bibinfo{booktitle}{\emph{Posters, Demos, Late-breaking Results and
  Workshop Proceedings of the 23rd Conference on User Modeling, Adaptation, and
  Personalization {(UMAP} 2015), Dublin, Ireland, June 29 - July 3, 2015}}
  \emph{(\bibinfo{series}{{CEUR} Workshop Proceedings},
  Vol.~\bibinfo{volume}{1388})},
  \bibfield{editor}{\bibinfo{person}{Alexandra~I. Cristea},
  \bibinfo{person}{Judith Masthoff}, \bibinfo{person}{Alan Said}, {and}
  \bibinfo{person}{Nava Tintarev}} (Eds.). \bibinfo{publisher}{CEUR-WS.org}.
\newblock


\bibitem[\protect\citeauthoryear{Guo, Tang, Ye, Li, and He}{Guo
  et~al\mbox{.}}{2017}]%
        {DBLP:conf/ijcai/GuoTYLH17}
\bibfield{author}{\bibinfo{person}{Huifeng Guo}, \bibinfo{person}{Ruiming
  Tang}, \bibinfo{person}{Yunming Ye}, \bibinfo{person}{Zhenguo Li}, {and}
  \bibinfo{person}{Xiuqiang He}.} \bibinfo{year}{2017}\natexlab{}.
\newblock \showarticletitle{DeepFM: {A} Factorization-Machine based Neural
  Network for {CTR} Prediction}. In \bibinfo{booktitle}{\emph{Proceedings of
  the Twenty-Sixth International Joint Conference on Artificial Intelligence,
  {IJCAI} 2017, Melbourne, Australia, August 19-25, 2017}},
  \bibfield{editor}{\bibinfo{person}{Carles Sierra}} (Ed.).
  \bibinfo{publisher}{ijcai.org}, \bibinfo{pages}{1725--1731}.
\newblock


\bibitem[\protect\citeauthoryear{Gupta, Hsia, Saraph, Wang, Reagen, Wei, Lee,
  Brooks, and Wu}{Gupta et~al\mbox{.}}{2020}]%
        {DBLP:conf/isca/GuptaHSWRWL0W20}
\bibfield{author}{\bibinfo{person}{Udit Gupta}, \bibinfo{person}{Samuel Hsia},
  \bibinfo{person}{Vikram Saraph}, \bibinfo{person}{Xiaodong Wang},
  \bibinfo{person}{Brandon Reagen}, \bibinfo{person}{Gu{-}Yeon Wei},
  \bibinfo{person}{Hsien{-}Hsin~S. Lee}, \bibinfo{person}{David Brooks}, {and}
  \bibinfo{person}{Carole{-}Jean Wu}.} \bibinfo{year}{2020}\natexlab{}.
\newblock \showarticletitle{DeepRecSys: {A} System for Optimizing End-To-End
  At-Scale Neural Recommendation Inference}. In \bibinfo{booktitle}{\emph{47th
  {ACM/IEEE} Annual International Symposium on Computer Architecture, {ISCA}
  2020, Valencia, Spain, May 30 - June 3, 2020}}. \bibinfo{publisher}{{IEEE}},
  \bibinfo{pages}{982--995}.
\newblock


\bibitem[\protect\citeauthoryear{He and McAuley}{He and McAuley}{2016}]%
        {DBLP:conf/aaai/HeM16}
\bibfield{author}{\bibinfo{person}{Ruining He} {and} \bibinfo{person}{Julian~J.
  McAuley}.} \bibinfo{year}{2016}\natexlab{}.
\newblock \showarticletitle{{VBPR:} Visual Bayesian Personalized Ranking from
  Implicit Feedback}. In \bibinfo{booktitle}{\emph{Proceedings of the Thirtieth
  {AAAI} Conference on Artificial Intelligence, February 12-17, 2016, Phoenix,
  Arizona, {USA}}}, \bibfield{editor}{\bibinfo{person}{Dale Schuurmans} {and}
  \bibinfo{person}{Michael~P. Wellman}} (Eds.). \bibinfo{publisher}{{AAAI}
  Press}, \bibinfo{pages}{144--150}.
\newblock


\bibitem[\protect\citeauthoryear{He and Chua}{He and Chua}{2017}]%
        {DBLP:conf/sigir/0001C17}
\bibfield{author}{\bibinfo{person}{Xiangnan He} {and}
  \bibinfo{person}{Tat{-}Seng Chua}.} \bibinfo{year}{2017}\natexlab{}.
\newblock \showarticletitle{Neural Factorization Machines for Sparse Predictive
  Analytics}. In \bibinfo{booktitle}{\emph{Proceedings of the 40th
  International {ACM} {SIGIR} Conference on Research and Development in
  Information Retrieval, Shinjuku, Tokyo, Japan, August 7-11, 2017}},
  \bibfield{editor}{\bibinfo{person}{Noriko Kando}, \bibinfo{person}{Tetsuya
  Sakai}, \bibinfo{person}{Hideo Joho}, \bibinfo{person}{Hang Li},
  \bibinfo{person}{Arjen~P. de~Vries}, {and} \bibinfo{person}{Ryen~W. White}}
  (Eds.). \bibinfo{publisher}{{ACM}}, \bibinfo{pages}{355--364}.
\newblock


\bibitem[\protect\citeauthoryear{He, Deng, Wang, Li, Zhang, and Wang}{He
  et~al\mbox{.}}{2020}]%
        {DBLP:conf/sigir/0001DWLZ020}
\bibfield{author}{\bibinfo{person}{Xiangnan He}, \bibinfo{person}{Kuan Deng},
  \bibinfo{person}{Xiang Wang}, \bibinfo{person}{Yan Li},
  \bibinfo{person}{Yong{-}Dong Zhang}, {and} \bibinfo{person}{Meng Wang}.}
  \bibinfo{year}{2020}\natexlab{}.
\newblock \showarticletitle{LightGCN: Simplifying and Powering Graph
  Convolution Network for Recommendation}. In
  \bibinfo{booktitle}{\emph{Proceedings of the 43rd International {ACM} {SIGIR}
  conference on research and development in Information Retrieval, {SIGIR}
  2020, Virtual Event, China, July 25-30, 2020}},
  \bibfield{editor}{\bibinfo{person}{Jimmy Huang}, \bibinfo{person}{Yi~Chang},
  \bibinfo{person}{Xueqi Cheng}, \bibinfo{person}{Jaap Kamps},
  \bibinfo{person}{Vanessa Murdock}, \bibinfo{person}{Ji{-}Rong Wen}, {and}
  \bibinfo{person}{Yiqun Liu}} (Eds.). \bibinfo{publisher}{{ACM}},
  \bibinfo{pages}{639--648}.
\newblock


\bibitem[\protect\citeauthoryear{He, Du, Wang, Tian, Tang, and Chua}{He
  et~al\mbox{.}}{2018a}]%
        {DBLP:conf/ijcai/0001DWTTC18}
\bibfield{author}{\bibinfo{person}{Xiangnan He}, \bibinfo{person}{Xiaoyu Du},
  \bibinfo{person}{Xiang Wang}, \bibinfo{person}{Feng Tian},
  \bibinfo{person}{Jinhui Tang}, {and} \bibinfo{person}{Tat{-}Seng Chua}.}
  \bibinfo{year}{2018}\natexlab{a}.
\newblock \showarticletitle{Outer Product-based Neural Collaborative
  Filtering}. In \bibinfo{booktitle}{\emph{Proceedings of the Twenty-Seventh
  International Joint Conference on Artificial Intelligence, {IJCAI} 2018, July
  13-19, 2018, Stockholm, Sweden}},
  \bibfield{editor}{\bibinfo{person}{J{\'{e}}r{\^{o}}me Lang}} (Ed.).
  \bibinfo{publisher}{ijcai.org}, \bibinfo{pages}{2227--2233}.
\newblock


\bibitem[\protect\citeauthoryear{He, He, Du, and Chua}{He
  et~al\mbox{.}}{2018b}]%
        {DBLP:conf/sigir/0001HDC18}
\bibfield{author}{\bibinfo{person}{Xiangnan He}, \bibinfo{person}{Zhankui He},
  \bibinfo{person}{Xiaoyu Du}, {and} \bibinfo{person}{Tat{-}Seng Chua}.}
  \bibinfo{year}{2018}\natexlab{b}.
\newblock \showarticletitle{Adversarial Personalized Ranking for
  Recommendation}. In \bibinfo{booktitle}{\emph{The 41st International {ACM}
  {SIGIR} Conference on Research {\&} Development in Information Retrieval,
  {SIGIR} 2018, Ann Arbor, MI, USA, July 08-12, 2018}},
  \bibfield{editor}{\bibinfo{person}{Kevyn Collins{-}Thompson},
  \bibinfo{person}{Qiaozhu Mei}, \bibinfo{person}{Brian~D. Davison},
  \bibinfo{person}{Yiqun Liu}, {and} \bibinfo{person}{Emine Yilmaz}} (Eds.).
  \bibinfo{publisher}{{ACM}}, \bibinfo{pages}{355--364}.
\newblock


\bibitem[\protect\citeauthoryear{He, He, Song, Liu, Jiang, and Chua}{He
  et~al\mbox{.}}{2018c}]%
        {DBLP:journals/tkde/HeHSLJC18}
\bibfield{author}{\bibinfo{person}{Xiangnan He}, \bibinfo{person}{Zhankui He},
  \bibinfo{person}{Jingkuan Song}, \bibinfo{person}{Zhenguang Liu},
  \bibinfo{person}{Yu{-}Gang Jiang}, {and} \bibinfo{person}{Tat{-}Seng Chua}.}
  \bibinfo{year}{2018}\natexlab{c}.
\newblock \showarticletitle{{NAIS:} Neural Attentive Item Similarity Model for
  Recommendation}.
\newblock \bibinfo{journal}{\emph{{IEEE} Trans. Knowl. Data Eng.}}
  \bibinfo{volume}{30}, \bibinfo{number}{12} (\bibinfo{year}{2018}),
  \bibinfo{pages}{2354--2366}.
\newblock


\bibitem[\protect\citeauthoryear{He, Liao, Zhang, Nie, Hu, and Chua}{He
  et~al\mbox{.}}{2017}]%
        {DBLP:conf/www/HeLZNHC17}
\bibfield{author}{\bibinfo{person}{Xiangnan He}, \bibinfo{person}{Lizi Liao},
  \bibinfo{person}{Hanwang Zhang}, \bibinfo{person}{Liqiang Nie},
  \bibinfo{person}{Xia Hu}, {and} \bibinfo{person}{Tat{-}Seng Chua}.}
  \bibinfo{year}{2017}\natexlab{}.
\newblock \showarticletitle{Neural Collaborative Filtering}. In
  \bibinfo{booktitle}{\emph{Proceedings of the 26th International Conference on
  World Wide Web, {WWW} 2017, Perth, Australia, April 3-7, 2017}},
  \bibfield{editor}{\bibinfo{person}{Rick Barrett}, \bibinfo{person}{Rick
  Cummings}, \bibinfo{person}{Eugene Agichtein}, {and} \bibinfo{person}{Evgeniy
  Gabrilovich}} (Eds.). \bibinfo{publisher}{{ACM}}, \bibinfo{pages}{173--182}.
\newblock


\bibitem[\protect\citeauthoryear{Hsieh, Yang, Cui, Lin, Belongie, and
  Estrin}{Hsieh et~al\mbox{.}}{2017}]%
        {DBLP:conf/www/HsiehYCLBE17}
\bibfield{author}{\bibinfo{person}{Cheng{-}Kang Hsieh}, \bibinfo{person}{Longqi
  Yang}, \bibinfo{person}{Yin Cui}, \bibinfo{person}{Tsung{-}Yi Lin},
  \bibinfo{person}{Serge~J. Belongie}, {and} \bibinfo{person}{Deborah Estrin}.}
  \bibinfo{year}{2017}\natexlab{}.
\newblock \showarticletitle{Collaborative Metric Learning}. In
  \bibinfo{booktitle}{\emph{Proceedings of the 26th International Conference on
  World Wide Web, {WWW} 2017, Perth, Australia, April 3-7, 2017}},
  \bibfield{editor}{\bibinfo{person}{Rick Barrett}, \bibinfo{person}{Rick
  Cummings}, \bibinfo{person}{Eugene Agichtein}, {and} \bibinfo{person}{Evgeniy
  Gabrilovich}} (Eds.). \bibinfo{publisher}{{ACM}}, \bibinfo{pages}{193--201}.
\newblock


\bibitem[\protect\citeauthoryear{Huang, He, Gao, Deng, Acero, and Heck}{Huang
  et~al\mbox{.}}{2013}]%
        {DBLP:conf/cikm/HuangHGDAH13}
\bibfield{author}{\bibinfo{person}{Po{-}Sen Huang}, \bibinfo{person}{Xiaodong
  He}, \bibinfo{person}{Jianfeng Gao}, \bibinfo{person}{Li Deng},
  \bibinfo{person}{Alex Acero}, {and} \bibinfo{person}{Larry~P. Heck}.}
  \bibinfo{year}{2013}\natexlab{}.
\newblock \showarticletitle{Learning deep structured semantic models for web
  search using clickthrough data}. In \bibinfo{booktitle}{\emph{22nd {ACM}
  International Conference on Information and Knowledge Management, CIKM'13,
  San Francisco, CA, USA, October 27 - November 1, 2013}},
  \bibfield{editor}{\bibinfo{person}{Qi~He}, \bibinfo{person}{Arun Iyengar},
  \bibinfo{person}{Wolfgang Nejdl}, \bibinfo{person}{Jian Pei}, {and}
  \bibinfo{person}{Rajeev Rastogi}} (Eds.). \bibinfo{publisher}{{ACM}},
  \bibinfo{pages}{2333--2338}.
\newblock


\bibitem[\protect\citeauthoryear{Hug}{Hug}{2020}]%
        {DBLP:journals/jossw/Hug20}
\bibfield{author}{\bibinfo{person}{Nicolas Hug}.}
  \bibinfo{year}{2020}\natexlab{}.
\newblock \showarticletitle{Surprise: {A} Python library for recommender
  systems}.
\newblock \bibinfo{journal}{\emph{J. Open Source Softw.}} \bibinfo{volume}{5},
  \bibinfo{number}{52} (\bibinfo{year}{2020}), \bibinfo{pages}{2174}.
\newblock


\bibitem[\protect\citeauthoryear{Hurley and Zhang}{Hurley and Zhang}{2011}]%
        {DBLP:journals/toit/HurleyZ11}
\bibfield{author}{\bibinfo{person}{Neil Hurley} {and} \bibinfo{person}{Mi
  Zhang}.} \bibinfo{year}{2011}\natexlab{}.
\newblock \showarticletitle{Novelty and Diversity in Top-N Recommendation -
  Analysis and Evaluation}.
\newblock \bibinfo{journal}{\emph{{ACM} Trans. Internet Techn.}}
  \bibinfo{volume}{10}, \bibinfo{number}{4} (\bibinfo{year}{2011}),
  \bibinfo{pages}{14:1--14:30}.
\newblock


\bibitem[\protect\citeauthoryear{Jamali and Ester}{Jamali and Ester}{2010}]%
        {DBLP:conf/recsys/JamaliE10}
\bibfield{author}{\bibinfo{person}{Mohsen Jamali} {and} \bibinfo{person}{Martin
  Ester}.} \bibinfo{year}{2010}\natexlab{}.
\newblock \showarticletitle{A matrix factorization technique with trust
  propagation for recommendation in social networks}. In
  \bibinfo{booktitle}{\emph{Proceedings of the 2010 {ACM} Conference on
  Recommender Systems, RecSys 2010, Barcelona, Spain, September 26-30, 2010}},
  \bibfield{editor}{\bibinfo{person}{Xavier Amatriain}, \bibinfo{person}{Marc
  Torrens}, \bibinfo{person}{Paul Resnick}, {and} \bibinfo{person}{Markus
  Zanker}} (Eds.). \bibinfo{publisher}{{ACM}}, \bibinfo{pages}{135--142}.
\newblock


\bibitem[\protect\citeauthoryear{Johnson}{Johnson}{2014}]%
        {johnson2014logistic}
\bibfield{author}{\bibinfo{person}{Christopher~C Johnson}.}
  \bibinfo{year}{2014}\natexlab{}.
\newblock \showarticletitle{Logistic matrix factorization for implicit feedback
  data}.
\newblock \bibinfo{journal}{\emph{Advances in Neural Information Processing
  Systems}} \bibinfo{volume}{27}, \bibinfo{number}{78} (\bibinfo{year}{2014}),
  \bibinfo{pages}{1--9}.
\newblock


\bibitem[\protect\citeauthoryear{Juan, Zhuang, Chin, and Lin}{Juan
  et~al\mbox{.}}{2016}]%
        {DBLP:conf/recsys/JuanZCL16}
\bibfield{author}{\bibinfo{person}{Yu{-}Chin Juan}, \bibinfo{person}{Yong
  Zhuang}, \bibinfo{person}{Wei{-}Sheng Chin}, {and}
  \bibinfo{person}{Chih{-}Jen Lin}.} \bibinfo{year}{2016}\natexlab{}.
\newblock \showarticletitle{Field-aware Factorization Machines for {CTR}
  Prediction}. In \bibinfo{booktitle}{\emph{Proceedings of the 10th {ACM}
  Conference on Recommender Systems, Boston, MA, USA, September 15-19, 2016}},
  \bibfield{editor}{\bibinfo{person}{Shilad Sen}, \bibinfo{person}{Werner
  Geyer}, \bibinfo{person}{Jill Freyne}, {and} \bibinfo{person}{Pablo
  Castells}} (Eds.). \bibinfo{publisher}{{ACM}}, \bibinfo{pages}{43--50}.
\newblock


\bibitem[\protect\citeauthoryear{Kabbur, Ning, and Karypis}{Kabbur
  et~al\mbox{.}}{2013}]%
        {DBLP:conf/kdd/KabburNK13}
\bibfield{author}{\bibinfo{person}{Santosh Kabbur}, \bibinfo{person}{Xia Ning},
  {and} \bibinfo{person}{George Karypis}.} \bibinfo{year}{2013}\natexlab{}.
\newblock \showarticletitle{{FISM:} factored item similarity models for top-N
  recommender systems}. In \bibinfo{booktitle}{\emph{The 19th {ACM} {SIGKDD}
  International Conference on Knowledge Discovery and Data Mining, {KDD} 2013,
  Chicago, IL, USA, August 11-14, 2013}},
  \bibfield{editor}{\bibinfo{person}{Inderjit~S. Dhillon},
  \bibinfo{person}{Yehuda Koren}, \bibinfo{person}{Rayid Ghani},
  \bibinfo{person}{Ted~E. Senator}, \bibinfo{person}{Paul Bradley},
  \bibinfo{person}{Rajesh Parekh}, \bibinfo{person}{Jingrui He},
  \bibinfo{person}{Robert~L. Grossman}, {and} \bibinfo{person}{Ramasamy
  Uthurusamy}} (Eds.). \bibinfo{publisher}{{ACM}}, \bibinfo{pages}{659--667}.
\newblock


\bibitem[\protect\citeauthoryear{Kang, Fang, Wang, and McAuley}{Kang
  et~al\mbox{.}}{2017}]%
        {DBLP:conf/icdm/KangFWM17}
\bibfield{author}{\bibinfo{person}{Wang{-}Cheng Kang}, \bibinfo{person}{Chen
  Fang}, \bibinfo{person}{Zhaowen Wang}, {and} \bibinfo{person}{Julian~J.
  McAuley}.} \bibinfo{year}{2017}\natexlab{}.
\newblock \showarticletitle{Visually-Aware Fashion Recommendation and Design
  with Generative Image Models}. In \bibinfo{booktitle}{\emph{2017 {IEEE}
  International Conference on Data Mining, {ICDM} 2017, New Orleans, LA, USA,
  November 18-21, 2017}}, \bibfield{editor}{\bibinfo{person}{Vijay Raghavan},
  \bibinfo{person}{Srinivas Aluru}, \bibinfo{person}{George Karypis},
  \bibinfo{person}{Lucio Miele}, {and} \bibinfo{person}{Xindong Wu}} (Eds.).
  \bibinfo{publisher}{{IEEE} Computer Society}, \bibinfo{pages}{207--216}.
\newblock


\bibitem[\protect\citeauthoryear{Kim, Park, Oh, Lee, and Yu}{Kim
  et~al\mbox{.}}{2016}]%
        {DBLP:conf/recsys/KimPOLY16}
\bibfield{author}{\bibinfo{person}{Dong~Hyun Kim}, \bibinfo{person}{Chanyoung
  Park}, \bibinfo{person}{Jinoh Oh}, \bibinfo{person}{Sungyoung Lee}, {and}
  \bibinfo{person}{Hwanjo Yu}.} \bibinfo{year}{2016}\natexlab{}.
\newblock \showarticletitle{Convolutional Matrix Factorization for Document
  Context-Aware Recommendation}. In \bibinfo{booktitle}{\emph{Proceedings of
  the 10th {ACM} Conference on Recommender Systems, Boston, MA, USA, September
  15-19, 2016}}, \bibfield{editor}{\bibinfo{person}{Shilad Sen},
  \bibinfo{person}{Werner Geyer}, \bibinfo{person}{Jill Freyne}, {and}
  \bibinfo{person}{Pablo Castells}} (Eds.). \bibinfo{publisher}{{ACM}},
  \bibinfo{pages}{233--240}.
\newblock


\bibitem[\protect\citeauthoryear{Konstan and Adomavicius}{Konstan and
  Adomavicius}{2013}]%
        {DBLP:conf/recsys/KonstanA13}
\bibfield{author}{\bibinfo{person}{Joseph~A. Konstan} {and}
  \bibinfo{person}{Gediminas Adomavicius}.} \bibinfo{year}{2013}\natexlab{}.
\newblock \showarticletitle{Toward identification and adoption of best
  practices in algorithmic recommender systems research}. In
  \bibinfo{booktitle}{\emph{Proceedings of the International Workshop on
  Reproducibility and Replication in Recommender Systems Evaluation, RepSys
  2013, Hong Kong, China, October 12, 2013}},
  \bibfield{editor}{\bibinfo{person}{Alejandro Bellog{\'{\i}}n},
  \bibinfo{person}{Pablo Castells}, \bibinfo{person}{Alan Said}, {and}
  \bibinfo{person}{Domonkos Tikk}} (Eds.). \bibinfo{publisher}{{ACM}},
  \bibinfo{pages}{23--28}.
\newblock


\bibitem[\protect\citeauthoryear{Koren}{Koren}{2008}]%
        {DBLP:conf/kdd/Koren08}
\bibfield{author}{\bibinfo{person}{Yehuda Koren}.}
  \bibinfo{year}{2008}\natexlab{}.
\newblock \showarticletitle{Factorization meets the neighborhood: a
  multifaceted collaborative filtering model}. In
  \bibinfo{booktitle}{\emph{Proceedings of the 14th {ACM} {SIGKDD}
  International Conference on Knowledge Discovery and Data Mining, Las Vegas,
  Nevada, USA, August 24-27, 2008}}, \bibfield{editor}{\bibinfo{person}{Ying
  Li}, \bibinfo{person}{Bing Liu}, {and} \bibinfo{person}{Sunita Sarawagi}}
  (Eds.). \bibinfo{publisher}{{ACM}}, \bibinfo{pages}{426--434}.
\newblock


\bibitem[\protect\citeauthoryear{Koren and Bell}{Koren and Bell}{2015}]%
        {DBLP:reference/sp/KorenB15}
\bibfield{author}{\bibinfo{person}{Yehuda Koren} {and}
  \bibinfo{person}{Robert~M. Bell}.} \bibinfo{year}{2015}\natexlab{}.
\newblock \showarticletitle{Advances in Collaborative Filtering}.
\newblock In \bibinfo{booktitle}{\emph{Recommender Systems Handbook}},
  \bibfield{editor}{\bibinfo{person}{Francesco Ricci}, \bibinfo{person}{Lior
  Rokach}, {and} \bibinfo{person}{Bracha Shapira}} (Eds.).
  \bibinfo{publisher}{Springer}, \bibinfo{pages}{77--118}.
\newblock


\bibitem[\protect\citeauthoryear{Koren, Bell, and Volinsky}{Koren
  et~al\mbox{.}}{2009}]%
        {DBLP:journals/computer/KorenBV09}
\bibfield{author}{\bibinfo{person}{Yehuda Koren}, \bibinfo{person}{Robert~M.
  Bell}, {and} \bibinfo{person}{Chris Volinsky}.}
  \bibinfo{year}{2009}\natexlab{}.
\newblock \showarticletitle{Matrix Factorization Techniques for Recommender
  Systems}.
\newblock \bibinfo{journal}{\emph{Computer}} \bibinfo{volume}{42},
  \bibinfo{number}{8} (\bibinfo{year}{2009}), \bibinfo{pages}{30--37}.
\newblock


\bibitem[\protect\citeauthoryear{Krestel, Fankhauser, and Nejdl}{Krestel
  et~al\mbox{.}}{2009}]%
        {DBLP:conf/recsys/KrestelFN09}
\bibfield{author}{\bibinfo{person}{Ralf Krestel}, \bibinfo{person}{Peter
  Fankhauser}, {and} \bibinfo{person}{Wolfgang Nejdl}.}
  \bibinfo{year}{2009}\natexlab{}.
\newblock \showarticletitle{Latent dirichlet allocation for tag
  recommendation}. In \bibinfo{booktitle}{\emph{Proceedings of the 2009 {ACM}
  Conference on Recommender Systems, RecSys 2009, New York, NY, USA, October
  23-25, 2009}}, \bibfield{editor}{\bibinfo{person}{Lawrence~D. Bergman},
  \bibinfo{person}{Alexander Tuzhilin}, \bibinfo{person}{Robin~D. Burke},
  \bibinfo{person}{Alexander Felfernig}, {and} \bibinfo{person}{Lars
  Schmidt{-}Thieme}} (Eds.). \bibinfo{publisher}{{ACM}},
  \bibinfo{pages}{61--68}.
\newblock


\bibitem[\protect\citeauthoryear{Krichene and Rendle}{Krichene and
  Rendle}{2020}]%
        {DBLP:conf/kdd/KricheneR20}
\bibfield{author}{\bibinfo{person}{Walid Krichene} {and}
  \bibinfo{person}{Steffen Rendle}.} \bibinfo{year}{2020}\natexlab{}.
\newblock \showarticletitle{On Sampled Metrics for Item Recommendation}. In
  \bibinfo{booktitle}{\emph{{KDD} '20: The 26th {ACM} {SIGKDD} Conference on
  Knowledge Discovery and Data Mining, Virtual Event, CA, USA, August 23-27,
  2020}}, \bibfield{editor}{\bibinfo{person}{Rajesh Gupta},
  \bibinfo{person}{Yan Liu}, \bibinfo{person}{Jiliang Tang}, {and}
  \bibinfo{person}{B.~Aditya Prakash}} (Eds.). \bibinfo{publisher}{{ACM}},
  \bibinfo{pages}{1748--1757}.
\newblock


\bibitem[\protect\citeauthoryear{Kula}{Kula}{2015}]%
        {DBLP:conf/recsys/Kula15}
\bibfield{author}{\bibinfo{person}{Maciej Kula}.}
  \bibinfo{year}{2015}\natexlab{}.
\newblock \showarticletitle{Metadata Embeddings for User and Item Cold-start
  Recommendations}. In \bibinfo{booktitle}{\emph{Proceedings of the 2nd
  Workshop on New Trends on Content-Based Recommender Systems co-located with
  9th {ACM} Conference on Recommender Systems (RecSys 2015), Vienna, Austria,
  September 16-20, 2015.}} \emph{(\bibinfo{series}{{CEUR} Workshop
  Proceedings}, Vol.~\bibinfo{volume}{1448})},
  \bibfield{editor}{\bibinfo{person}{Toine Bogers} {and}
  \bibinfo{person}{Marijn Koolen}} (Eds.). \bibinfo{publisher}{CEUR-WS.org},
  \bibinfo{pages}{14--21}.
\newblock


\bibitem[\protect\citeauthoryear{Kula}{Kula}{2017}]%
        {kula2017spotlight}
\bibfield{author}{\bibinfo{person}{Maciej Kula}.}
  \bibinfo{year}{2017}\natexlab{}.
\newblock \bibinfo{title}{Spotlight}.
\newblock
  \bibinfo{howpublished}{\url{https://github.com/maciejkula/spotlight}}.
\newblock


\bibitem[\protect\citeauthoryear{Lemire and Maclachlan}{Lemire and
  Maclachlan}{2005}]%
        {DBLP:conf/sdm/LemireM05}
\bibfield{author}{\bibinfo{person}{Daniel Lemire} {and} \bibinfo{person}{Anna
  Maclachlan}.} \bibinfo{year}{2005}\natexlab{}.
\newblock \showarticletitle{Slope One Predictors for Online Rating-Based
  Collaborative Filtering}. In \bibinfo{booktitle}{\emph{Proceedings of the
  2005 {SIAM} International Conference on Data Mining, {SDM} 2005, Newport
  Beach, CA, USA, April 21-23, 2005}},
  \bibfield{editor}{\bibinfo{person}{Hillol Kargupta}, \bibinfo{person}{Jaideep
  Srivastava}, \bibinfo{person}{Chandrika Kamath}, {and}
  \bibinfo{person}{Arnold Goodman}} (Eds.). \bibinfo{publisher}{{SIAM}},
  \bibinfo{pages}{471--475}.
\newblock


\bibitem[\protect\citeauthoryear{Li, Jin, Gao, and Liu}{Li
  et~al\mbox{.}}{2020}]%
        {DBLP:conf/kdd/LiJGL20}
\bibfield{author}{\bibinfo{person}{Dong Li}, \bibinfo{person}{Ruoming Jin},
  \bibinfo{person}{Jing Gao}, {and} \bibinfo{person}{Zhi Liu}.}
  \bibinfo{year}{2020}\natexlab{}.
\newblock \showarticletitle{On Sampling Top-K Recommendation Evaluation}. In
  \bibinfo{booktitle}{\emph{{KDD} '20: The 26th {ACM} {SIGKDD} Conference on
  Knowledge Discovery and Data Mining, Virtual Event, CA, USA, August 23-27,
  2020}}, \bibfield{editor}{\bibinfo{person}{Rajesh Gupta},
  \bibinfo{person}{Yan Liu}, \bibinfo{person}{Jiliang Tang}, {and}
  \bibinfo{person}{B.~Aditya Prakash}} (Eds.). \bibinfo{publisher}{{ACM}},
  \bibinfo{pages}{2114--2124}.
\newblock


\bibitem[\protect\citeauthoryear{Liang, Krishnan, Hoffman, and Jebara}{Liang
  et~al\mbox{.}}{2018}]%
        {DBLP:conf/www/LiangKHJ18}
\bibfield{author}{\bibinfo{person}{Dawen Liang}, \bibinfo{person}{Rahul~G.
  Krishnan}, \bibinfo{person}{Matthew~D. Hoffman}, {and} \bibinfo{person}{Tony
  Jebara}.} \bibinfo{year}{2018}\natexlab{}.
\newblock \showarticletitle{Variational Autoencoders for Collaborative
  Filtering}. In \bibinfo{booktitle}{\emph{Proceedings of the 2018 World Wide
  Web Conference on World Wide Web, {WWW} 2018, Lyon, France, April 23-27,
  2018}}, \bibfield{editor}{\bibinfo{person}{Pierre{-}Antoine Champin},
  \bibinfo{person}{Fabien~L. Gandon}, \bibinfo{person}{Mounia Lalmas}, {and}
  \bibinfo{person}{Panagiotis~G. Ipeirotis}} (Eds.).
  \bibinfo{publisher}{{ACM}}, \bibinfo{pages}{689--698}.
\newblock


\bibitem[\protect\citeauthoryear{Linden, Smith, and York}{Linden
  et~al\mbox{.}}{2003}]%
        {DBLP:journals/internet/LindenSY03}
\bibfield{author}{\bibinfo{person}{Greg Linden}, \bibinfo{person}{Brent Smith},
  {and} \bibinfo{person}{Jeremy York}.} \bibinfo{year}{2003}\natexlab{}.
\newblock \showarticletitle{Amazon.com Recommendations: Item-to-Item
  Collaborative Filtering}.
\newblock \bibinfo{journal}{\emph{{IEEE} Internet Comput.}}
  \bibinfo{volume}{7}, \bibinfo{number}{1} (\bibinfo{year}{2003}),
  \bibinfo{pages}{76--80}.
\newblock


\bibitem[\protect\citeauthoryear{Liu, Wu, and Wang}{Liu et~al\mbox{.}}{2017}]%
        {DBLP:conf/sigir/LiuWW17}
\bibfield{author}{\bibinfo{person}{Qiang Liu}, \bibinfo{person}{Shu Wu}, {and}
  \bibinfo{person}{Liang Wang}.} \bibinfo{year}{2017}\natexlab{}.
\newblock \showarticletitle{DeepStyle: Learning User Preferences for Visual
  Recommendation}. In \bibinfo{booktitle}{\emph{Proceedings of the 40th
  International {ACM} {SIGIR} Conference on Research and Development in
  Information Retrieval, Shinjuku, Tokyo, Japan, August 7-11, 2017}},
  \bibfield{editor}{\bibinfo{person}{Noriko Kando}, \bibinfo{person}{Tetsuya
  Sakai}, \bibinfo{person}{Hideo Joho}, \bibinfo{person}{Hang Li},
  \bibinfo{person}{Arjen~P. de~Vries}, {and} \bibinfo{person}{Ryen~W. White}}
  (Eds.). \bibinfo{publisher}{{ACM}}, \bibinfo{pages}{841--844}.
\newblock


\bibitem[\protect\citeauthoryear{Ludewig, Mauro, Latifi, and Jannach}{Ludewig
  et~al\mbox{.}}{2019}]%
        {DBLP:conf/recsys/LudewigMLJ19}
\bibfield{author}{\bibinfo{person}{Malte Ludewig}, \bibinfo{person}{Noemi
  Mauro}, \bibinfo{person}{Sara Latifi}, {and} \bibinfo{person}{Dietmar
  Jannach}.} \bibinfo{year}{2019}\natexlab{}.
\newblock \showarticletitle{Performance comparison of neural and non-neural
  approaches to session-based recommendation}. In
  \bibinfo{booktitle}{\emph{Proceedings of the 13th {ACM} Conference on
  Recommender Systems, RecSys 2019, Copenhagen, Denmark, September 16-20,
  2019}}, \bibfield{editor}{\bibinfo{person}{Toine Bogers},
  \bibinfo{person}{Alan Said}, \bibinfo{person}{Peter Brusilovsky}, {and}
  \bibinfo{person}{Domonkos Tikk}} (Eds.). \bibinfo{publisher}{{ACM}},
  \bibinfo{pages}{462--466}.
\newblock


\bibitem[\protect\citeauthoryear{Ludewig, Mauro, Latifi, and Jannach}{Ludewig
  et~al\mbox{.}}{2021}]%
        {DBLP:journals/umuai/LudewigMLJ21}
\bibfield{author}{\bibinfo{person}{Malte Ludewig}, \bibinfo{person}{Noemi
  Mauro}, \bibinfo{person}{Sara Latifi}, {and} \bibinfo{person}{Dietmar
  Jannach}.} \bibinfo{year}{2021}\natexlab{}.
\newblock \showarticletitle{Empirical analysis of session-based recommendation
  algorithms}.
\newblock \bibinfo{journal}{\emph{User Model. User Adapt. Interact.}}
  \bibinfo{volume}{31}, \bibinfo{number}{1} (\bibinfo{year}{2021}),
  \bibinfo{pages}{149--181}.
\newblock


\bibitem[\protect\citeauthoryear{Luo, Zhou, Xia, and Zhu}{Luo
  et~al\mbox{.}}{2014}]%
        {DBLP:journals/tii/LuoZXZ14}
\bibfield{author}{\bibinfo{person}{Xin Luo}, \bibinfo{person}{Mengchu Zhou},
  \bibinfo{person}{Yunni Xia}, {and} \bibinfo{person}{Qingsheng Zhu}.}
  \bibinfo{year}{2014}\natexlab{}.
\newblock \showarticletitle{An Efficient Non-Negative
  Matrix-Factorization-Based Approach to Collaborative Filtering for
  Recommender Systems}.
\newblock \bibinfo{journal}{\emph{{IEEE} Trans. Ind. Informatics}}
  \bibinfo{volume}{10}, \bibinfo{number}{2} (\bibinfo{year}{2014}),
  \bibinfo{pages}{1273--1284}.
\newblock


\bibitem[\protect\citeauthoryear{Ma, Yang, Lyu, and King}{Ma
  et~al\mbox{.}}{2008}]%
        {DBLP:conf/cikm/MaYLK08a}
\bibfield{author}{\bibinfo{person}{Hao Ma}, \bibinfo{person}{Haixuan Yang},
  \bibinfo{person}{Michael~R. Lyu}, {and} \bibinfo{person}{Irwin King}.}
  \bibinfo{year}{2008}\natexlab{}.
\newblock \showarticletitle{SoRec: social recommendation using probabilistic
  matrix factorization}. In \bibinfo{booktitle}{\emph{Proceedings of the 17th
  {ACM} Conference on Information and Knowledge Management, {CIKM} 2008, Napa
  Valley, California, USA, October 26-30, 2008}},
  \bibfield{editor}{\bibinfo{person}{James~G. Shanahan}, \bibinfo{person}{Sihem
  Amer{-}Yahia}, \bibinfo{person}{Ioana Manolescu}, \bibinfo{person}{Yi~Zhang},
  \bibinfo{person}{David~A. Evans}, \bibinfo{person}{Aleksander Kolcz},
  \bibinfo{person}{Key{-}Sun Choi}, {and} \bibinfo{person}{Abdur Chowdhury}}
  (Eds.). \bibinfo{publisher}{{ACM}}, \bibinfo{pages}{931--940}.
\newblock


\bibitem[\protect\citeauthoryear{Ma, Zhou, Liu, Lyu, and King}{Ma
  et~al\mbox{.}}{2011}]%
        {DBLP:conf/wsdm/MaZLLK11}
\bibfield{author}{\bibinfo{person}{Hao Ma}, \bibinfo{person}{Dengyong Zhou},
  \bibinfo{person}{Chao Liu}, \bibinfo{person}{Michael~R. Lyu}, {and}
  \bibinfo{person}{Irwin King}.} \bibinfo{year}{2011}\natexlab{}.
\newblock \showarticletitle{Recommender systems with social regularization}. In
  \bibinfo{booktitle}{\emph{Proceedings of the Forth International Conference
  on Web Search and Web Data Mining, {WSDM} 2011, Hong Kong, China, February
  9-12, 2011}}, \bibfield{editor}{\bibinfo{person}{Irwin King},
  \bibinfo{person}{Wolfgang Nejdl}, {and} \bibinfo{person}{Hang Li}} (Eds.).
  \bibinfo{publisher}{{ACM}}, \bibinfo{pages}{287--296}.
\newblock


\bibitem[\protect\citeauthoryear{McNee, Riedl, and Konstan}{McNee
  et~al\mbox{.}}{2006}]%
        {DBLP:conf/chi/McNeeRK06}
\bibfield{author}{\bibinfo{person}{Sean~M. McNee}, \bibinfo{person}{John
  Riedl}, {and} \bibinfo{person}{Joseph~A. Konstan}.}
  \bibinfo{year}{2006}\natexlab{}.
\newblock \showarticletitle{Being accurate is not enough: how accuracy metrics
  have hurt recommender systems}. In \bibinfo{booktitle}{\emph{Extended
  Abstracts Proceedings of the 2006 Conference on Human Factors in Computing
  Systems, {CHI} 2006, Montr{\'{e}}al, Qu{\'{e}}bec, Canada, April 22-27,
  2006}}, \bibfield{editor}{\bibinfo{person}{Gary~M. Olson} {and}
  \bibinfo{person}{Robin Jeffries}} (Eds.). \bibinfo{publisher}{{ACM}},
  \bibinfo{pages}{1097--1101}.
\newblock


\bibitem[\protect\citeauthoryear{Meng, McCreadie, Macdonald, and Ounis}{Meng
  et~al\mbox{.}}{2020}]%
        {DBLP:conf/recsys/MengMMO20}
\bibfield{author}{\bibinfo{person}{Zaiqiao Meng}, \bibinfo{person}{Richard
  McCreadie}, \bibinfo{person}{Craig Macdonald}, {and} \bibinfo{person}{Iadh
  Ounis}.} \bibinfo{year}{2020}\natexlab{}.
\newblock \showarticletitle{Exploring Data Splitting Strategies for the
  Evaluation of Recommendation Models}. In \bibinfo{booktitle}{\emph{RecSys
  2020: Fourteenth {ACM} Conference on Recommender Systems, Virtual Event,
  Brazil, September 22-26, 2020}}, \bibfield{editor}{\bibinfo{person}{Rodrygo
  L.~T. Santos}, \bibinfo{person}{Leandro~Balby Marinho},
  \bibinfo{person}{Elizabeth~M. Daly}, \bibinfo{person}{Li~Chen},
  \bibinfo{person}{Kim Falk}, \bibinfo{person}{Noam Koenigstein}, {and}
  \bibinfo{person}{Edleno~Silva de~Moura}} (Eds.). \bibinfo{publisher}{{ACM}},
  \bibinfo{pages}{681--686}.
\newblock


\bibitem[\protect\citeauthoryear{Ning and Karypis}{Ning and Karypis}{2011}]%
        {DBLP:conf/icdm/NingK11}
\bibfield{author}{\bibinfo{person}{Xia Ning} {and} \bibinfo{person}{George
  Karypis}.} \bibinfo{year}{2011}\natexlab{}.
\newblock \showarticletitle{{SLIM:} Sparse Linear Methods for Top-N Recommender
  Systems}. In \bibinfo{booktitle}{\emph{11th {IEEE} International Conference
  on Data Mining, {ICDM} 2011, Vancouver, BC, Canada, December 11-14, 2011}},
  \bibfield{editor}{\bibinfo{person}{Diane~J. Cook}, \bibinfo{person}{Jian
  Pei}, \bibinfo{person}{Wei Wang}, \bibinfo{person}{Osmar~R. Za{\"{\i}}ane},
  {and} \bibinfo{person}{Xindong Wu}} (Eds.). \bibinfo{publisher}{{IEEE}
  Computer Society}, \bibinfo{pages}{497--506}.
\newblock


\bibitem[\protect\citeauthoryear{Niu, Caverlee, and Lu}{Niu
  et~al\mbox{.}}{2018}]%
        {DBLP:conf/wsdm/NiuCL18}
\bibfield{author}{\bibinfo{person}{Wei Niu}, \bibinfo{person}{James Caverlee},
  {and} \bibinfo{person}{Haokai Lu}.} \bibinfo{year}{2018}\natexlab{}.
\newblock \showarticletitle{Neural Personalized Ranking for Image
  Recommendation}. In \bibinfo{booktitle}{\emph{Proceedings of the Eleventh
  {ACM} International Conference on Web Search and Data Mining, {WSDM} 2018,
  Marina Del Rey, CA, USA, February 5-9, 2018}},
  \bibfield{editor}{\bibinfo{person}{Yi~Chang}, \bibinfo{person}{Chengxiang
  Zhai}, \bibinfo{person}{Yan Liu}, {and} \bibinfo{person}{Yoelle Maarek}}
  (Eds.). \bibinfo{publisher}{{ACM}}, \bibinfo{pages}{423--431}.
\newblock


\bibitem[\protect\citeauthoryear{Noia, Mirizzi, Ostuni, Romito, and
  Zanker}{Noia et~al\mbox{.}}{2012}]%
        {DBLP:conf/i-semantics/NoiaMORZ12}
\bibfield{author}{\bibinfo{person}{Tommaso~Di Noia}, \bibinfo{person}{Roberto
  Mirizzi}, \bibinfo{person}{Vito~Claudio Ostuni}, \bibinfo{person}{Davide
  Romito}, {and} \bibinfo{person}{Markus Zanker}.}
  \bibinfo{year}{2012}\natexlab{}.
\newblock \showarticletitle{Linked open data to support content-based
  recommender systems}. In \bibinfo{booktitle}{\emph{{I-SEMANTICS} 2012 - 8th
  International Conference on Semantic Systems, {I-SEMANTICS} '12, Graz,
  Austria, September 5-7, 2012}}, \bibfield{editor}{\bibinfo{person}{Valentina
  Presutti} {and} \bibinfo{person}{Helena~Sofia Pinto}} (Eds.).
  \bibinfo{publisher}{{ACM}}, \bibinfo{pages}{1--8}.
\newblock


\bibitem[\protect\citeauthoryear{Rendle}{Rendle}{2010}]%
        {DBLP:conf/icdm/Rendle10}
\bibfield{author}{\bibinfo{person}{Steffen Rendle}.}
  \bibinfo{year}{2010}\natexlab{}.
\newblock \showarticletitle{Factorization Machines}. In
  \bibinfo{booktitle}{\emph{{ICDM} 2010, The 10th {IEEE} International
  Conference on Data Mining, Sydney, Australia, 14-17 December 2010}},
  \bibfield{editor}{\bibinfo{person}{Geoffrey~I. Webb}, \bibinfo{person}{Bing
  Liu}, \bibinfo{person}{Chengqi Zhang}, \bibinfo{person}{Dimitrios Gunopulos},
  {and} \bibinfo{person}{Xindong Wu}} (Eds.). \bibinfo{publisher}{{IEEE}
  Computer Society}, \bibinfo{pages}{995--1000}.
\newblock


\bibitem[\protect\citeauthoryear{Rendle, Freudenthaler, Gantner, and
  Schmidt{-}Thieme}{Rendle et~al\mbox{.}}{2009}]%
        {DBLP:conf/uai/RendleFGS09}
\bibfield{author}{\bibinfo{person}{Steffen Rendle}, \bibinfo{person}{Christoph
  Freudenthaler}, \bibinfo{person}{Zeno Gantner}, {and} \bibinfo{person}{Lars
  Schmidt{-}Thieme}.} \bibinfo{year}{2009}\natexlab{}.
\newblock \showarticletitle{{BPR:} Bayesian Personalized Ranking from Implicit
  Feedback}. In \bibinfo{booktitle}{\emph{{UAI} 2009, Proceedings of the
  Twenty-Fifth Conference on Uncertainty in Artificial Intelligence, Montreal,
  QC, Canada, June 18-21, 2009}}, \bibfield{editor}{\bibinfo{person}{Jeff~A.
  Bilmes} {and} \bibinfo{person}{Andrew~Y. Ng}} (Eds.).
  \bibinfo{publisher}{{AUAI} Press}, \bibinfo{pages}{452--461}.
\newblock


\bibitem[\protect\citeauthoryear{Rendle, Krichene, Zhang, and Anderson}{Rendle
  et~al\mbox{.}}{2020}]%
        {DBLP:conf/recsys/RendleKZA20}
\bibfield{author}{\bibinfo{person}{Steffen Rendle}, \bibinfo{person}{Walid
  Krichene}, \bibinfo{person}{Li Zhang}, {and} \bibinfo{person}{John~R.
  Anderson}.} \bibinfo{year}{2020}\natexlab{}.
\newblock \showarticletitle{Neural Collaborative Filtering vs. Matrix
  Factorization Revisited}. In \bibinfo{booktitle}{\emph{RecSys 2020:
  Fourteenth {ACM} Conference on Recommender Systems, Virtual Event, Brazil,
  September 22-26, 2020}}, \bibfield{editor}{\bibinfo{person}{Rodrygo L.~T.
  Santos}, \bibinfo{person}{Leandro~Balby Marinho},
  \bibinfo{person}{Elizabeth~M. Daly}, \bibinfo{person}{Li~Chen},
  \bibinfo{person}{Kim Falk}, \bibinfo{person}{Noam Koenigstein}, {and}
  \bibinfo{person}{Edleno~Silva de~Moura}} (Eds.). \bibinfo{publisher}{{ACM}},
  \bibinfo{pages}{240--248}.
\newblock


\bibitem[\protect\citeauthoryear{Rendle, Zhang, and Koren}{Rendle
  et~al\mbox{.}}{2019}]%
        {DBLP:journals/corr/abs-1905-01395}
\bibfield{author}{\bibinfo{person}{Steffen Rendle}, \bibinfo{person}{Li Zhang},
  {and} \bibinfo{person}{Yehuda Koren}.} \bibinfo{year}{2019}\natexlab{}.
\newblock \showarticletitle{On the Difficulty of Evaluating Baselines: {A}
  Study on Recommender Systems}.
\newblock \bibinfo{journal}{\emph{CoRR}}  \bibinfo{volume}{abs/1905.01395}
  (\bibinfo{year}{2019}).
\newblock


\bibitem[\protect\citeauthoryear{Resnick, Iacovou, Suchak, Bergstrom, and
  Riedl}{Resnick et~al\mbox{.}}{1994}]%
        {DBLP:conf/cscw/ResnickISBR94}
\bibfield{author}{\bibinfo{person}{Paul Resnick}, \bibinfo{person}{Neophytos
  Iacovou}, \bibinfo{person}{Mitesh Suchak}, \bibinfo{person}{Peter Bergstrom},
  {and} \bibinfo{person}{John Riedl}.} \bibinfo{year}{1994}\natexlab{}.
\newblock \showarticletitle{GroupLens: An Open Architecture for Collaborative
  Filtering of Netnews}. In \bibinfo{booktitle}{\emph{{CSCW} '94, Proceedings
  of the Conference on Computer Supported Cooperative Work, Chapel Hill, NC,
  USA, October 22-26, 1994}}, \bibfield{editor}{\bibinfo{person}{John~B.
  Smith}, \bibinfo{person}{F.~Donelson Smith}, {and} \bibinfo{person}{Thomas~W.
  Malone}} (Eds.). \bibinfo{publisher}{{ACM}}, \bibinfo{pages}{175--186}.
\newblock


\bibitem[\protect\citeauthoryear{Said and Bellog{\'{\i}}n}{Said and
  Bellog{\'{\i}}n}{2014a}]%
        {DBLP:conf/recsys/SaidB14}
\bibfield{author}{\bibinfo{person}{Alan Said} {and} \bibinfo{person}{Alejandro
  Bellog{\'{\i}}n}.} \bibinfo{year}{2014}\natexlab{a}.
\newblock \showarticletitle{Comparative recommender system evaluation:
  benchmarking recommendation frameworks}. In \bibinfo{booktitle}{\emph{Eighth
  {ACM} Conference on Recommender Systems, RecSys '14, Foster City, Silicon
  Valley, CA, {USA} - October 06 - 10, 2014}},
  \bibfield{editor}{\bibinfo{person}{Alfred Kobsa},
  \bibinfo{person}{Michelle~X. Zhou}, \bibinfo{person}{Martin Ester}, {and}
  \bibinfo{person}{Yehuda Koren}} (Eds.). \bibinfo{publisher}{{ACM}},
  \bibinfo{pages}{129--136}.
\newblock


\bibitem[\protect\citeauthoryear{Said and Bellog{\'{\i}}n}{Said and
  Bellog{\'{\i}}n}{2014b}]%
        {DBLP:conf/recsys/SaidB14a}
\bibfield{author}{\bibinfo{person}{Alan Said} {and} \bibinfo{person}{Alejandro
  Bellog{\'{\i}}n}.} \bibinfo{year}{2014}\natexlab{b}.
\newblock \showarticletitle{Rival: a toolkit to foster reproducibility in
  recommender system evaluation}. In \bibinfo{booktitle}{\emph{Eighth {ACM}
  Conference on Recommender Systems, RecSys '14, Foster City, Silicon Valley,
  CA, {USA} - October 06 - 10, 2014}},
  \bibfield{editor}{\bibinfo{person}{Alfred Kobsa},
  \bibinfo{person}{Michelle~X. Zhou}, \bibinfo{person}{Martin Ester}, {and}
  \bibinfo{person}{Yehuda Koren}} (Eds.). \bibinfo{publisher}{{ACM}},
  \bibinfo{pages}{371--372}.
\newblock


\bibitem[\protect\citeauthoryear{Salah, Truong, and Lauw}{Salah
  et~al\mbox{.}}{2020}]%
        {DBLP:journals/jmlr/SalahTL20}
\bibfield{author}{\bibinfo{person}{Aghiles Salah}, \bibinfo{person}{Quoc{-}Tuan
  Truong}, {and} \bibinfo{person}{Hady~W. Lauw}.}
  \bibinfo{year}{2020}\natexlab{}.
\newblock \showarticletitle{Cornac: {A} Comparative Framework for Multimodal
  Recommender Systems}.
\newblock \bibinfo{journal}{\emph{J. Mach. Learn. Res.}}  \bibinfo{volume}{21}
  (\bibinfo{year}{2020}), \bibinfo{pages}{95:1--95:5}.
\newblock


\bibitem[\protect\citeauthoryear{Salakhutdinov and Mnih}{Salakhutdinov and
  Mnih}{2007}]%
        {DBLP:conf/nips/SalakhutdinovM07}
\bibfield{author}{\bibinfo{person}{Ruslan Salakhutdinov} {and}
  \bibinfo{person}{Andriy Mnih}.} \bibinfo{year}{2007}\natexlab{}.
\newblock \showarticletitle{Probabilistic Matrix Factorization}. In
  \bibinfo{booktitle}{\emph{Advances in Neural Information Processing Systems
  20, Proceedings of the Twenty-First Annual Conference on Neural Information
  Processing Systems, Vancouver, British Columbia, Canada, December 3-6,
  2007}}, \bibfield{editor}{\bibinfo{person}{John~C. Platt},
  \bibinfo{person}{Daphne Koller}, \bibinfo{person}{Yoram Singer}, {and}
  \bibinfo{person}{Sam~T. Roweis}} (Eds.). \bibinfo{publisher}{Curran
  Associates, Inc.}, \bibinfo{pages}{1257--1264}.
\newblock


\bibitem[\protect\citeauthoryear{Sarwar, Karypis, Konstan, and Riedl}{Sarwar
  et~al\mbox{.}}{2001}]%
        {DBLP:conf/www/SarwarKKR01}
\bibfield{author}{\bibinfo{person}{Badrul~Munir Sarwar},
  \bibinfo{person}{George Karypis}, \bibinfo{person}{Joseph~A. Konstan}, {and}
  \bibinfo{person}{John Riedl}.} \bibinfo{year}{2001}\natexlab{}.
\newblock \showarticletitle{Item-based collaborative filtering recommendation
  algorithms}. In \bibinfo{booktitle}{\emph{Proceedings of the Tenth
  International World Wide Web Conference, {WWW} 10, Hong Kong, China, May 1-5,
  2001}}, \bibfield{editor}{\bibinfo{person}{Vincent~Y. Shen},
  \bibinfo{person}{Nobuo Saito}, \bibinfo{person}{Michael~R. Lyu}, {and}
  \bibinfo{person}{Mary~Ellen Zurko}} (Eds.). \bibinfo{publisher}{{ACM}},
  \bibinfo{pages}{285--295}.
\newblock


\bibitem[\protect\citeauthoryear{Schröder, Thiele, and Lehner}{Schröder
  et~al\mbox{.}}{2011}]%
        {Schroder201178}
\bibfield{author}{\bibinfo{person}{G. Schröder}, \bibinfo{person}{M. Thiele},
  {and} \bibinfo{person}{W. Lehner}.} \bibinfo{year}{2011}\natexlab{}.
\newblock \showarticletitle{Setting goals and choosing metrics for recommender
  system evaluations}.
\newblock \bibinfo{journal}{\emph{CEUR Workshop Proceedings}}
  \bibinfo{volume}{811} (\bibinfo{year}{2011}), \bibinfo{pages}{78--85}.
\newblock


\bibitem[\protect\citeauthoryear{Sedhain, Menon, Sanner, and Xie}{Sedhain
  et~al\mbox{.}}{2015}]%
        {DBLP:conf/www/SedhainMSX15}
\bibfield{author}{\bibinfo{person}{Suvash Sedhain},
  \bibinfo{person}{Aditya~Krishna Menon}, \bibinfo{person}{Scott Sanner}, {and}
  \bibinfo{person}{Lexing Xie}.} \bibinfo{year}{2015}\natexlab{}.
\newblock \showarticletitle{AutoRec: Autoencoders Meet Collaborative
  Filtering}. In \bibinfo{booktitle}{\emph{Proceedings of the 24th
  International Conference on World Wide Web Companion, {WWW} 2015, Florence,
  Italy, May 18-22, 2015 - Companion Volume}},
  \bibfield{editor}{\bibinfo{person}{Aldo Gangemi}, \bibinfo{person}{Stefano
  Leonardi}, {and} \bibinfo{person}{Alessandro Panconesi}} (Eds.).
  \bibinfo{publisher}{{ACM}}, \bibinfo{pages}{111--112}.
\newblock


\bibitem[\protect\citeauthoryear{Sun, Yu, Fang, Yang, Qu, Zhang, and Geng}{Sun
  et~al\mbox{.}}{2020}]%
        {DBLP:conf/recsys/SunY00Q0G20}
\bibfield{author}{\bibinfo{person}{Zhu Sun}, \bibinfo{person}{Di Yu},
  \bibinfo{person}{Hui Fang}, \bibinfo{person}{Jie Yang},
  \bibinfo{person}{Xinghua Qu}, \bibinfo{person}{Jie Zhang}, {and}
  \bibinfo{person}{Cong Geng}.} \bibinfo{year}{2020}\natexlab{}.
\newblock \showarticletitle{Are We Evaluating Rigorously? Benchmarking
  Recommendation for Reproducible Evaluation and Fair Comparison}. In
  \bibinfo{booktitle}{\emph{RecSys 2020: Fourteenth {ACM} Conference on
  Recommender Systems, Virtual Event, Brazil, September 22-26, 2020}},
  \bibfield{editor}{\bibinfo{person}{Rodrygo L.~T. Santos},
  \bibinfo{person}{Leandro~Balby Marinho}, \bibinfo{person}{Elizabeth~M. Daly},
  \bibinfo{person}{Li~Chen}, \bibinfo{person}{Kim Falk}, \bibinfo{person}{Noam
  Koenigstein}, {and} \bibinfo{person}{Edleno~Silva de~Moura}} (Eds.).
  \bibinfo{publisher}{{ACM}}, \bibinfo{pages}{23--32}.
\newblock


\bibitem[\protect\citeauthoryear{Tang, Du, He, Yuan, Tian, and Chua}{Tang
  et~al\mbox{.}}{2020}]%
        {DBLP:journals/tkde/TangDHYTC20}
\bibfield{author}{\bibinfo{person}{Jinhui Tang}, \bibinfo{person}{Xiaoyu Du},
  \bibinfo{person}{Xiangnan He}, \bibinfo{person}{Fajie Yuan},
  \bibinfo{person}{Qi Tian}, {and} \bibinfo{person}{Tat{-}Seng Chua}.}
  \bibinfo{year}{2020}\natexlab{}.
\newblock \showarticletitle{Adversarial Training Towards Robust Multimedia
  Recommender System}.
\newblock \bibinfo{journal}{\emph{{IEEE} Trans. Knowl. Data Eng.}}
  \bibinfo{volume}{32}, \bibinfo{number}{5} (\bibinfo{year}{2020}),
  \bibinfo{pages}{855--867}.
\newblock


\bibitem[\protect\citeauthoryear{Tang and Wang}{Tang and Wang}{2018}]%
        {DBLP:conf/wsdm/TangW18}
\bibfield{author}{\bibinfo{person}{Jiaxi Tang} {and} \bibinfo{person}{Ke
  Wang}.} \bibinfo{year}{2018}\natexlab{}.
\newblock \showarticletitle{Personalized Top-N Sequential Recommendation via
  Convolutional Sequence Embedding}. In \bibinfo{booktitle}{\emph{Proceedings
  of the Eleventh {ACM} International Conference on Web Search and Data Mining,
  {WSDM} 2018, Marina Del Rey, CA, USA, February 5-9, 2018}},
  \bibfield{editor}{\bibinfo{person}{Yi~Chang}, \bibinfo{person}{Chengxiang
  Zhai}, \bibinfo{person}{Yan Liu}, {and} \bibinfo{person}{Yoelle Maarek}}
  (Eds.). \bibinfo{publisher}{{ACM}}, \bibinfo{pages}{565--573}.
\newblock


\bibitem[\protect\citeauthoryear{Tsintzou, Pitoura, and Tsaparas}{Tsintzou
  et~al\mbox{.}}{2019}]%
        {DBLP:conf/recsys/TsintzouPT19}
\bibfield{author}{\bibinfo{person}{Virginia Tsintzou},
  \bibinfo{person}{Evaggelia Pitoura}, {and} \bibinfo{person}{Panayiotis
  Tsaparas}.} \bibinfo{year}{2019}\natexlab{}.
\newblock \showarticletitle{Bias Disparity in Recommendation Systems}. In
  \bibinfo{booktitle}{\emph{Proceedings of the Workshop on Recommendation in
  Multi-stakeholder Environments co-located with the 13th {ACM} Conference on
  Recommender Systems (RecSys 2019), Copenhagen, Denmark, September 20, 2019}}
  \emph{(\bibinfo{series}{{CEUR} Workshop Proceedings},
  Vol.~\bibinfo{volume}{2440})}, \bibfield{editor}{\bibinfo{person}{Robin
  Burke}, \bibinfo{person}{Himan Abdollahpouri}, \bibinfo{person}{Edward~C.
  Malthouse}, \bibinfo{person}{K.~P. Thai}, {and} \bibinfo{person}{Yongfeng
  Zhang}} (Eds.). \bibinfo{publisher}{CEUR-WS.org}.
\newblock


\bibitem[\protect\citeauthoryear{Valcarce, Bellog{\'{\i}}n, Parapar, and
  Castells}{Valcarce et~al\mbox{.}}{2018}]%
        {DBLP:conf/recsys/ValcarceBPC18}
\bibfield{author}{\bibinfo{person}{Daniel Valcarce}, \bibinfo{person}{Alejandro
  Bellog{\'{\i}}n}, \bibinfo{person}{Javier Parapar}, {and}
  \bibinfo{person}{Pablo Castells}.} \bibinfo{year}{2018}\natexlab{}.
\newblock \showarticletitle{On the robustness and discriminative power of
  information retrieval metrics for top-N recommendation}. In
  \bibinfo{booktitle}{\emph{Proceedings of the 12th {ACM} Conference on
  Recommender Systems, RecSys 2018, Vancouver, BC, Canada, October 2-7, 2018}},
  \bibfield{editor}{\bibinfo{person}{Sole Pera}, \bibinfo{person}{Michael~D.
  Ekstrand}, \bibinfo{person}{Xavier Amatriain}, {and} \bibinfo{person}{John
  O'Donovan}} (Eds.). \bibinfo{publisher}{{ACM}}, \bibinfo{pages}{260--268}.
\newblock


\bibitem[\protect\citeauthoryear{Vargas}{Vargas}{2014}]%
        {DBLP:conf/sigir/Vargas14}
\bibfield{author}{\bibinfo{person}{Sa{\'{u}}l Vargas}.}
  \bibinfo{year}{2014}\natexlab{}.
\newblock \showarticletitle{Novelty and diversity enhancement and evaluation in
  recommender systems and information retrieval}. In
  \bibinfo{booktitle}{\emph{The 37th International {ACM} {SIGIR} Conference on
  Research and Development in Information Retrieval, {SIGIR} '14, Gold Coast ,
  QLD, Australia - July 06 - 11, 2014}},
  \bibfield{editor}{\bibinfo{person}{Shlomo Geva}, \bibinfo{person}{Andrew
  Trotman}, \bibinfo{person}{Peter Bruza}, \bibinfo{person}{Charles L.~A.
  Clarke}, {and} \bibinfo{person}{Kalervo J{\"{a}}rvelin}} (Eds.).
  \bibinfo{publisher}{{ACM}}, \bibinfo{pages}{1281}.
\newblock


\bibitem[\protect\citeauthoryear{Vargas and Castells}{Vargas and
  Castells}{2011}]%
        {DBLP:conf/recsys/VargasC11}
\bibfield{author}{\bibinfo{person}{Saul Vargas} {and} \bibinfo{person}{Pablo
  Castells}.} \bibinfo{year}{2011}\natexlab{}.
\newblock \showarticletitle{Rank and relevance in novelty and diversity metrics
  for recommender systems}. In \bibinfo{booktitle}{\emph{Proceedings of the
  2011 {ACM} Conference on Recommender Systems, RecSys 2011, Chicago, IL, USA,
  October 23-27, 2011}}, \bibfield{editor}{\bibinfo{person}{Bamshad Mobasher},
  \bibinfo{person}{Robin~D. Burke}, \bibinfo{person}{Dietmar Jannach}, {and}
  \bibinfo{person}{Gediminas Adomavicius}} (Eds.). \bibinfo{publisher}{{ACM}},
  \bibinfo{pages}{109--116}.
\newblock


\bibitem[\protect\citeauthoryear{Wang, Yu, Zhang, Gong, Xu, Wang, Zhang, and
  Zhang}{Wang et~al\mbox{.}}{2017}]%
        {DBLP:conf/sigir/WangYZGXWZZ17}
\bibfield{author}{\bibinfo{person}{Jun Wang}, \bibinfo{person}{Lantao Yu},
  \bibinfo{person}{Weinan Zhang}, \bibinfo{person}{Yu Gong},
  \bibinfo{person}{Yinghui Xu}, \bibinfo{person}{Benyou Wang},
  \bibinfo{person}{Peng Zhang}, {and} \bibinfo{person}{Dell Zhang}.}
  \bibinfo{year}{2017}\natexlab{}.
\newblock \showarticletitle{{IRGAN:} {A} Minimax Game for Unifying Generative
  and Discriminative Information Retrieval Models}. In
  \bibinfo{booktitle}{\emph{Proceedings of the 40th International {ACM} {SIGIR}
  Conference on Research and Development in Information Retrieval, Shinjuku,
  Tokyo, Japan, August 7-11, 2017}}, \bibfield{editor}{\bibinfo{person}{Noriko
  Kando}, \bibinfo{person}{Tetsuya Sakai}, \bibinfo{person}{Hideo Joho},
  \bibinfo{person}{Hang Li}, \bibinfo{person}{Arjen~P. de~Vries}, {and}
  \bibinfo{person}{Ryen~W. White}} (Eds.). \bibinfo{publisher}{{ACM}},
  \bibinfo{pages}{515--524}.
\newblock


\bibitem[\protect\citeauthoryear{Wang, He, Wang, Feng, and Chua}{Wang
  et~al\mbox{.}}{2019}]%
        {DBLP:conf/sigir/Wang0WFC19}
\bibfield{author}{\bibinfo{person}{Xiang Wang}, \bibinfo{person}{Xiangnan He},
  \bibinfo{person}{Meng Wang}, \bibinfo{person}{Fuli Feng}, {and}
  \bibinfo{person}{Tat{-}Seng Chua}.} \bibinfo{year}{2019}\natexlab{}.
\newblock \showarticletitle{Neural Graph Collaborative Filtering}. In
  \bibinfo{booktitle}{\emph{Proceedings of the 42nd International {ACM} {SIGIR}
  Conference on Research and Development in Information Retrieval, {SIGIR}
  2019, Paris, France, July 21-25, 2019}},
  \bibfield{editor}{\bibinfo{person}{Benjamin Piwowarski}, \bibinfo{person}{Max
  Chevalier}, \bibinfo{person}{{\'{E}}ric Gaussier}, \bibinfo{person}{Yoelle
  Maarek}, \bibinfo{person}{Jian{-}Yun Nie}, {and} \bibinfo{person}{Falk
  Scholer}} (Eds.). \bibinfo{publisher}{{ACM}}, \bibinfo{pages}{165--174}.
\newblock


\bibitem[\protect\citeauthoryear{Wu, DuBois, Zheng, and Ester}{Wu
  et~al\mbox{.}}{2016}]%
        {DBLP:conf/wsdm/WuDZE16}
\bibfield{author}{\bibinfo{person}{Yao Wu}, \bibinfo{person}{Christopher
  DuBois}, \bibinfo{person}{Alice~X. Zheng}, {and} \bibinfo{person}{Martin
  Ester}.} \bibinfo{year}{2016}\natexlab{}.
\newblock \showarticletitle{Collaborative Denoising Auto-Encoders for Top-N
  Recommender Systems}. In \bibinfo{booktitle}{\emph{Proceedings of the Ninth
  {ACM} International Conference on Web Search and Data Mining, San Francisco,
  CA, USA, February 22-25, 2016}}, \bibfield{editor}{\bibinfo{person}{Paul~N.
  Bennett}, \bibinfo{person}{Vanja Josifovski}, \bibinfo{person}{Jennifer
  Neville}, {and} \bibinfo{person}{Filip Radlinski}} (Eds.).
  \bibinfo{publisher}{{ACM}}, \bibinfo{pages}{153--162}.
\newblock


\bibitem[\protect\citeauthoryear{Xiao, Ye, He, Zhang, Wu, and Chua}{Xiao
  et~al\mbox{.}}{2017}]%
        {DBLP:conf/ijcai/XiaoY0ZWC17}
\bibfield{author}{\bibinfo{person}{Jun Xiao}, \bibinfo{person}{Hao Ye},
  \bibinfo{person}{Xiangnan He}, \bibinfo{person}{Hanwang Zhang},
  \bibinfo{person}{Fei Wu}, {and} \bibinfo{person}{Tat{-}Seng Chua}.}
  \bibinfo{year}{2017}\natexlab{}.
\newblock \showarticletitle{Attentional Factorization Machines: Learning the
  Weight of Feature Interactions via Attention Networks}. In
  \bibinfo{booktitle}{\emph{Proceedings of the Twenty-Sixth International Joint
  Conference on Artificial Intelligence, {IJCAI} 2017, Melbourne, Australia,
  August 19-25, 2017}}, \bibfield{editor}{\bibinfo{person}{Carles Sierra}}
  (Ed.). \bibinfo{publisher}{ijcai.org}, \bibinfo{pages}{3119--3125}.
\newblock


\bibitem[\protect\citeauthoryear{Xue, Dai, Zhang, Huang, and Chen}{Xue
  et~al\mbox{.}}{2017}]%
        {DBLP:conf/ijcai/XueDZHC17}
\bibfield{author}{\bibinfo{person}{Hong{-}Jian Xue}, \bibinfo{person}{Xinyu
  Dai}, \bibinfo{person}{Jianbing Zhang}, \bibinfo{person}{Shujian Huang},
  {and} \bibinfo{person}{Jiajun Chen}.} \bibinfo{year}{2017}\natexlab{}.
\newblock \showarticletitle{Deep Matrix Factorization Models for Recommender
  Systems}. In \bibinfo{booktitle}{\emph{Proceedings of the Twenty-Sixth
  International Joint Conference on Artificial Intelligence, {IJCAI} 2017,
  Melbourne, Australia, August 19-25, 2017}},
  \bibfield{editor}{\bibinfo{person}{Carles Sierra}} (Ed.).
  \bibinfo{publisher}{ijcai.org}, \bibinfo{pages}{3203--3209}.
\newblock


\bibitem[\protect\citeauthoryear{Yang, Bagdasaryan, Gruenstein, Hsieh, and
  Estrin}{Yang et~al\mbox{.}}{2018}]%
        {DBLP:conf/wsdm/YangBGHE18}
\bibfield{author}{\bibinfo{person}{Longqi Yang}, \bibinfo{person}{Eugene
  Bagdasaryan}, \bibinfo{person}{Joshua Gruenstein},
  \bibinfo{person}{Cheng{-}Kang Hsieh}, {and} \bibinfo{person}{Deborah
  Estrin}.} \bibinfo{year}{2018}\natexlab{}.
\newblock \showarticletitle{OpenRec: {A} Modular Framework for Extensible and
  Adaptable Recommendation Algorithms}. In
  \bibinfo{booktitle}{\emph{Proceedings of the Eleventh {ACM} International
  Conference on Web Search and Data Mining, {WSDM} 2018, Marina Del Rey, CA,
  USA, February 5-9, 2018}}, \bibfield{editor}{\bibinfo{person}{Yi~Chang},
  \bibinfo{person}{Chengxiang Zhai}, \bibinfo{person}{Yan Liu}, {and}
  \bibinfo{person}{Yoelle Maarek}} (Eds.). \bibinfo{publisher}{{ACM}},
  \bibinfo{pages}{664--672}.
\newblock


\bibitem[\protect\citeauthoryear{Yin, Cui, Li, Yao, and Chen}{Yin
  et~al\mbox{.}}{2012}]%
        {DBLP:journals/pvldb/YinCLYC12}
\bibfield{author}{\bibinfo{person}{Hongzhi Yin}, \bibinfo{person}{Bin Cui},
  \bibinfo{person}{Jing Li}, \bibinfo{person}{Junjie Yao}, {and}
  \bibinfo{person}{Chen Chen}.} \bibinfo{year}{2012}\natexlab{}.
\newblock \showarticletitle{Challenging the Long Tail Recommendation}.
\newblock \bibinfo{journal}{\emph{Proc. {VLDB} Endow.}} \bibinfo{volume}{5},
  \bibinfo{number}{9} (\bibinfo{year}{2012}), \bibinfo{pages}{896--907}.
\newblock


\bibitem[\protect\citeauthoryear{Yu, Gao, Yin, Li, Gao, and Wang}{Yu
  et~al\mbox{.}}{2019}]%
        {DBLP:conf/icdm/Yu0YLGW19}
\bibfield{author}{\bibinfo{person}{Junliang Yu}, \bibinfo{person}{Min Gao},
  \bibinfo{person}{Hongzhi Yin}, \bibinfo{person}{Jundong Li},
  \bibinfo{person}{Chongming Gao}, {and} \bibinfo{person}{Qinyong Wang}.}
  \bibinfo{year}{2019}\natexlab{}.
\newblock \showarticletitle{Generating Reliable Friends via Adversarial
  Training to Improve Social Recommendation}. In \bibinfo{booktitle}{\emph{2019
  {IEEE} International Conference on Data Mining, {ICDM} 2019, Beijing, China,
  November 8-11, 2019}}, \bibfield{editor}{\bibinfo{person}{Jianyong Wang},
  \bibinfo{person}{Kyuseok Shim}, {and} \bibinfo{person}{Xindong Wu}} (Eds.).
  \bibinfo{publisher}{{IEEE}}, \bibinfo{pages}{768--777}.
\newblock


\bibitem[\protect\citeauthoryear{Zhai, Cohen, and Lafferty}{Zhai
  et~al\mbox{.}}{2003}]%
        {DBLP:conf/sigir/ZhaiCL03}
\bibfield{author}{\bibinfo{person}{ChengXiang Zhai},
  \bibinfo{person}{William~W. Cohen}, {and} \bibinfo{person}{John~D.
  Lafferty}.} \bibinfo{year}{2003}\natexlab{}.
\newblock \showarticletitle{Beyond independent relevance: methods and
  evaluation metrics for subtopic retrieval}. In
  \bibinfo{booktitle}{\emph{{SIGIR} 2003: Proceedings of the 26th Annual
  International {ACM} {SIGIR} Conference on Research and Development in
  Information Retrieval, July 28 - August 1, 2003, Toronto, Canada}},
  \bibfield{editor}{\bibinfo{person}{Charles L.~A. Clarke},
  \bibinfo{person}{Gordon~V. Cormack}, \bibinfo{person}{Jamie Callan},
  \bibinfo{person}{David Hawking}, {and} \bibinfo{person}{Alan~F. Smeaton}}
  (Eds.). \bibinfo{publisher}{{ACM}}, \bibinfo{pages}{10--17}.
\newblock


\bibitem[\protect\citeauthoryear{Zhang, Wang, Ford, and Makedon}{Zhang
  et~al\mbox{.}}{2006}]%
        {DBLP:conf/sdm/ZhangWFM06}
\bibfield{author}{\bibinfo{person}{Sheng Zhang}, \bibinfo{person}{Weihong
  Wang}, \bibinfo{person}{James Ford}, {and} \bibinfo{person}{Fillia Makedon}.}
  \bibinfo{year}{2006}\natexlab{}.
\newblock \showarticletitle{Learning from Incomplete Ratings Using Non-negative
  Matrix Factorization}. In \bibinfo{booktitle}{\emph{Proceedings of the Sixth
  {SIAM} International Conference on Data Mining, April 20-22, 2006, Bethesda,
  MD, {USA}}}, \bibfield{editor}{\bibinfo{person}{Joydeep Ghosh},
  \bibinfo{person}{Diane Lambert}, \bibinfo{person}{David~B. Skillicorn}, {and}
  \bibinfo{person}{Jaideep Srivastava}} (Eds.). \bibinfo{publisher}{{SIAM}},
  \bibinfo{pages}{549--553}.
\newblock


\bibitem[\protect\citeauthoryear{Zhao, Mu, Hou, Lin, Li, Chen, Lu, Wang, Tian,
  Pan, Min, Feng, Fan, Chen, Wang, Ji, Li, Wang, and Wen}{Zhao
  et~al\mbox{.}}{2020}]%
        {DBLP:journals/corr/abs-2011-01731}
\bibfield{author}{\bibinfo{person}{Wayne~Xin Zhao}, \bibinfo{person}{Shanlei
  Mu}, \bibinfo{person}{Yupeng Hou}, \bibinfo{person}{Zihan Lin},
  \bibinfo{person}{Kaiyuan Li}, \bibinfo{person}{Yushuo Chen},
  \bibinfo{person}{Yujie Lu}, \bibinfo{person}{Hui Wang},
  \bibinfo{person}{Changxin Tian}, \bibinfo{person}{Xingyu Pan},
  \bibinfo{person}{Yingqian Min}, \bibinfo{person}{Zhichao Feng},
  \bibinfo{person}{Xinyan Fan}, \bibinfo{person}{Xu Chen},
  \bibinfo{person}{Pengfei Wang}, \bibinfo{person}{Wendi Ji},
  \bibinfo{person}{Yaliang Li}, \bibinfo{person}{Xiaoling Wang}, {and}
  \bibinfo{person}{Ji{-}Rong Wen}.} \bibinfo{year}{2020}\natexlab{}.
\newblock \showarticletitle{RecBole: Towards a Unified, Comprehensive and
  Efficient Framework for Recommendation Algorithms}.
\newblock \bibinfo{journal}{\emph{CoRR}}  \bibinfo{volume}{abs/2011.01731}
  (\bibinfo{year}{2020}).
\newblock


\bibitem[\protect\citeauthoryear{Zhou, Zhu, Song, Fan, Zhu, Ma, Yan, Jin, Li,
  and Gai}{Zhou et~al\mbox{.}}{2018}]%
        {DBLP:conf/kdd/ZhouZSFZMYJLG18}
\bibfield{author}{\bibinfo{person}{Guorui Zhou}, \bibinfo{person}{Xiaoqiang
  Zhu}, \bibinfo{person}{Chengru Song}, \bibinfo{person}{Ying Fan},
  \bibinfo{person}{Han Zhu}, \bibinfo{person}{Xiao Ma},
  \bibinfo{person}{Yanghui Yan}, \bibinfo{person}{Junqi Jin},
  \bibinfo{person}{Han Li}, {and} \bibinfo{person}{Kun Gai}.}
  \bibinfo{year}{2018}\natexlab{}.
\newblock \showarticletitle{Deep Interest Network for Click-Through Rate
  Prediction}. In \bibinfo{booktitle}{\emph{Proceedings of the 24th {ACM}
  {SIGKDD} International Conference on Knowledge Discovery {\&} Data Mining,
  {KDD} 2018, London, UK, August 19-23, 2018}},
  \bibfield{editor}{\bibinfo{person}{Yike Guo} {and} \bibinfo{person}{Faisal
  Farooq}} (Eds.). \bibinfo{publisher}{{ACM}}, \bibinfo{pages}{1059--1068}.
\newblock


\bibitem[\protect\citeauthoryear{Zhu, He, Zhao, Zhang, Wang, and Caverlee}{Zhu
  et~al\mbox{.}}{2021}]%
        {Zhu21}
\bibfield{author}{\bibinfo{person}{Ziwei Zhu}, \bibinfo{person}{Yun He},
  \bibinfo{person}{Xing Zhao}, \bibinfo{person}{Yin Zhang},
  \bibinfo{person}{Jianling Wang}, {and} \bibinfo{person}{James Caverlee}.}
  \bibinfo{year}{2021}\natexlab{}.
\newblock \showarticletitle{Popularity-Opportunity Bias in Collaborative
  Filtering}. In \bibinfo{booktitle}{\emph{Proceedings of the Fourteenth ACM
  International Conference on Web Search and Data Mining (WSDM '21), March
  8–12, 2021, Virtual Event, Israel}}. \bibinfo{publisher}{{ACM}}.
\newblock


\bibitem[\protect\citeauthoryear{Zhu, Hu, and Caverlee}{Zhu
  et~al\mbox{.}}{2018}]%
        {DBLP:conf/cikm/ZhuHC18}
\bibfield{author}{\bibinfo{person}{Ziwei Zhu}, \bibinfo{person}{Xia Hu}, {and}
  \bibinfo{person}{James Caverlee}.} \bibinfo{year}{2018}\natexlab{}.
\newblock \showarticletitle{Fairness-Aware Tensor-Based Recommendation}. In
  \bibinfo{booktitle}{\emph{Proceedings of the 27th {ACM} International
  Conference on Information and Knowledge Management, {CIKM} 2018, Torino,
  Italy, October 22-26, 2018}}, \bibfield{editor}{\bibinfo{person}{Alfredo
  Cuzzocrea}, \bibinfo{person}{James Allan}, \bibinfo{person}{Norman~W. Paton},
  \bibinfo{person}{Divesh Srivastava}, \bibinfo{person}{Rakesh Agrawal},
  \bibinfo{person}{Andrei~Z. Broder}, \bibinfo{person}{Mohammed~J. Zaki},
  \bibinfo{person}{K.~Sel{\c{c}}uk Candan}, \bibinfo{person}{Alexandros
  Labrinidis}, \bibinfo{person}{Assaf Schuster}, {and} \bibinfo{person}{Haixun
  Wang}} (Eds.). \bibinfo{publisher}{{ACM}}, \bibinfo{pages}{1153--1162}.
\newblock


\bibitem[\protect\citeauthoryear{Zhu, Wang, and Caverlee}{Zhu
  et~al\mbox{.}}{2020}]%
        {DBLP:conf/sigir/ZhuWC20}
\bibfield{author}{\bibinfo{person}{Ziwei Zhu}, \bibinfo{person}{Jianling Wang},
  {and} \bibinfo{person}{James Caverlee}.} \bibinfo{year}{2020}\natexlab{}.
\newblock \showarticletitle{Measuring and Mitigating Item Under-Recommendation
  Bias in Personalized Ranking Systems}. In
  \bibinfo{booktitle}{\emph{Proceedings of the 43rd International {ACM} {SIGIR}
  conference on research and development in Information Retrieval, {SIGIR}
  2020, Virtual Event, China, July 25-30, 2020}},
  \bibfield{editor}{\bibinfo{person}{Jimmy Huang}, \bibinfo{person}{Yi~Chang},
  \bibinfo{person}{Xueqi Cheng}, \bibinfo{person}{Jaap Kamps},
  \bibinfo{person}{Vanessa Murdock}, \bibinfo{person}{Ji{-}Rong Wen}, {and}
  \bibinfo{person}{Yiqun Liu}} (Eds.). \bibinfo{publisher}{{ACM}},
  \bibinfo{pages}{449--458}.
\newblock


\end{thebibliography}

\end{document}